\documentclass[usenatbib]{mn2e}
\usepackage{graphicx}
\usepackage{color}
\title[Uncertainties in  the SFR calibrations]
{Uncertainties in the calibrations of star formation rate}

\author[F. Zhang, L. Li, Y. Kang, Y. Zhuang, Z. Han]
{Fenghui~Zhang\thanks{E-mail: zhangfh@ynao.ac.cn; zhang\_fh@hotmail.com}$^{1,2}$, Lifang~Li$^{1,2}$, Xiaoyu~Kang$^{1,2,3}$ Yulong~Zhuang$^{1,2,3}$ and Zhanwen~Han$^{1,2}$\\
$^1$National Astronomical Observatories/Yunnan Observatory, Chinese Academy of Sciences, Kunming, 650011, China \\
$^2$Key Laboratory for the Structure and Evolution of Celestial Objects, Chinese Academy of Sciences, Kunming, 650011, China \\
$^3$Graduate University of the Chinese Academy of Science, Beijing 100049, China}

\font\hf = cmsl7 scaled \magstep 0

\begin{document}
\date{\today}
\pagerange{\pageref{firstpage}--\pageref{lastpage}}
\pubyear{2011}
\maketitle
\label{firstpage}

\begin{abstract}
The calibrations of star formation rate (SFR) are prone to be affected by many factors, such as metallicity, initial mass function (IMF), evolutionary population synthesis (EPS) models and so on. In this paper we will discuss the effects of binary interactions, metallicity, EPS models and IMF on several widely used SFR calibrations based on the EPS models of Yunnan with and without binary interactions, BC03, {\hf SB99}, {\hf P\'EGASE} and POPSTAR.
The inclusion of binary interactions makes these SFR conversion coefficients smaller (less than 0.2\,dex), and these differences increase with metallicity.
The differences in the calibration coefficient between SFR and the luminosity of $\rm H\alpha$ recombination line (C$_{\rm H\alpha}$) and that between SFR and the ultraviolet (UV) fluxes at 1500 and 2800\,$\rm \AA$ (C$_{i, {\rm UV}}$), caused by IMF, are independent of metallicity (0.03-0.33\,dex) except $\Delta$C$_{\rm H\alpha, IMF}$ when using the POPSTAR and $\Delta$C$_{i, {\rm UV, IMF}}$ when using the {\hf P\'EGASE} models.
Moreover, we find that $L_{\rm 2800}$ is not suitable to the linear calibration of SFR at low metallicities.

At last, we compare the effects of these several factors on the SFR calibrations considered in this paper. The effects of metallicity/IMF and EPS models on the C$_{\rm H\alpha}$ and C$_{\rm FIR}$ (the conversion coefficient between SFR and the far-infrared flux) are the largest among these factors, respectively.
For the calibration between SFR and C$_{i, {\rm UV}}$, the effects of these several factors are comparable. 
\end{abstract}


\section{Introduction}
Star formation rate (SFR) is an important parameter in the studies of galaxy formation and evolution. The luminosity of H$\alpha$ recombination line ($L_{\rm H\alpha}$), the luminosity of [OII]$\lambda$3727 forbidden line doublet ($L_{\rm [OII]}$), the ultraviolet (UV, $L_{i, {\rm UV}}$) and far-infrared (FIR, $L_{\rm FIR}$) continuum fluxes are the commonly used traces of star formation rate (SFR, \citealt[][hearafter K98]{ken98}). HCN \citep{gao04}, radio luminosity \citep{hop03} and X-ray luminosity have even been used as SFR indicators.

In this work we only study the first four indicators (i.e. $L_{\rm H\alpha}$, $L_{\rm [OII]}$, $L_{i, {\rm UV}}$ and $L_{\rm FIR}$). In history, the calibrations of SFR in terms of these diagnostics are often obtained at solar metallicity and by using the evolutionary population synthesis (EPS) models without binary interactions, however, we know that binary systems are common in the Universe and the stars are not always at solar metallicity.

Binary stars are common in the Universe. Upwards of 50\% of field stars are in binary systems. In young massive stellar populations (SPs), the binary fraction is close to one (\citealt{kou07,kob07}, also the references from \citealt{eld12}). Moreover, in the Tarantula Nebula, the binary frequency among massive stars is high, with the ESO's VLT-FLAMES Tarantula Survey (VFTS) establishing that approximately two out of three massive stars are born in a binary system that will interact during their evolution (from \citealt{cro12}).

\citet{zha04} have included binary interactions in the EPS models. Now, more and more studies began to pay an attention to the effect of binary interactions.
\citet{her11} have considered binary interactions in their EPS models. \citet{san09} have investigated the impact of binary-star yields on the spectra of galaxies.
\citet{kan12} have considered binary interactions in the colour and chemical evolutions of M33.
\citet{zhy12} have analyzed the differences between the model and observed spectra of globular clusters by using the EPS models comprising binaries.
\citet{zha09} have investigated the effect of binary interactions on the determination of photometric redshift for galaxies. \citet{eld12} and \citet[][Paper I, at solar metallicity]{zha12} have investigated the effects of massive binaries and binaries on the SFR calibrations, respectively.
Moreover, \citet{hur05} have included binary interactions in the {\hf Nbody4} code \citep{ara99}. \cite{spu99} and \citet{and12} have included binaries in the {\hf Nbody6++} and {\hf STARLAB} codes, respectively.
At last, some researchers have investigated the effect of binary interactions on the observations \citep{deg08}.

Besides the effects of metallicity and binary interactions, these SFR calibrations are prone to be affected by initial mass function (IMF), EPS models and stellar rotation and so on.
K98 has even summarized that the effects of metallicity and IMF on the SFR($L_{\rm UV}$) and SFR($L_{\rm H\alpha}$) calibration factors reach to $\sim$0.3 and $\sim$0.1\, dex, but in this work we will see that these effects are underestimated.
\citet{hor13}, \cite{lei08} and \citet{mey00} have even studied the effect of stellar rotation on the SFR calibrations, it can cause to 30 and 40 per cent of the differences in the SFR($L_{\rm UV}$) and SFR($L_{\rm H\alpha}$) calibration factors.

Motivated by the above mentioned facts, in this paper we will present these calibrations of SFR at non-solar metallicities by using several sets of EPS models and discuss the effects of binary interactions, metallicity, EPS models and IMF on these SFR calibrations.
This work is also helpful to check consistency among SFRs, which are obtained by using different indicators, and to understand the galaxy properties (such as dust attenuation). Several studies have compared the SFRs derived from different indicators and have concluded that these SFRs agree broadly with each other. However, we can see that there exists discrepancy (for example, see Fig.\,1 of \citealt{hop04}). The mismatch among SFRs obtained from different indicators would lead to misunderstand the properties of galaxies. From this study we will see that the discrepancy between the SFR obtained by using $L_{\rm H\alpha}$ and $L_{\rm UV}$ diagnostics would be enlarged if using the SFR calibration relations at low metallicities or those in the case of considering binary interactions.

The outline of the paper is as follows. In Section 2 we describe the used EPS models and algorithms. In section 3 we discuss some results concerning binary evolutions. In Section 4 we present the conversion coefficients between SFR and these tracers and discuss the effects of binary interactions and metallicity on these SFR calibrations based on the Yunnan EPS models. In Section 5 we present the conversion coefficients between SFR and these traces by using the other EPS models, compare the conclusions (the effect of metallicity on these SFR calibrations) with those from the Yunnan models and discuss the effects of EPS models, IMF and metallicity on these SFR calibrations. In Section 6 we summary the effects of binary interactions, metallicity, IMF and EPS models on these SFR calibrations and discuss the influences of metallicity and binary interactions on the discrepancy in the SFR between derived from $L_{\rm H\alpha}$ and $L_{\rm UV}$ indicators. Finally we present a summary and conclusions in Section 7.

\section{Models and algorithms}
\begin{table*}
\centering
\caption{Definition of models (the first column) and description of the used EPS models [including the name, the IMFs, the upper and lower mass limits ($M_{\rm l}$, $M_{\rm u}$)  and the metallicities, from the second to the last columns].}
\begin{tabular}{lll ccc ccc cc}
\hline
  Model     & \multicolumn{4}{c}{EPS models} \\
 \hline
                & name & IMF & $M_{\rm l},M_{\rm u}$ (${\rm M_{\odot}}$)& metallicity {\color{red} \bf ($Z$)}\\
\hline
 A/B  & Yunnan        & MS79      & 0.10, 100 & 0.0001/0.0003/0.001/0.004/0.01/0.02/0.03   \\
 C-S55/Cha03  & BC03           & S55/Cha03 & 0.10, 100 & {\color{red}0.0001}/{\color{red}0.0004}/0.004/0.008/0.02/{\color{red}0.05}/- - -        \\
 D-S55/K93'  & {\hf SB99}     & S55/K93'  & 0.10, 100 & - - - - /{\color{red} 0.0004}/0.004/0.008/0.02/{\color{red} 0.05}/- - -      \\
 E-S55/K93'  & {\hf P\'EGASE} & S55/K93'  & 0.10, 100 & {\color{red}0.0001}/{\color{red}0.0004}/0.004/0.008/0.02/{\color{red}0.05}/{\color{red}0.10}  \\
 F-S55'/K01  & POPSTAR        & S55'/K01   & 0.15, 100 & {\color{red}0.0001}/{\color{red}0.0004}/0.004/0.008/0.02/{\color{red}0.05}/- - -      \\
\hline
\end{tabular}
\label{Tab:mod-des}
\end{table*}

In order to present the SFR calibrations in terms of $L_{\rm H\alpha}$, $L_{\rm [OII]}$, $L_{\rm 1500}$, $L_{\rm 2800}$ and $L_{\rm FIR}$ for each set of models at different metallicities, we need to present these parameters for various types of galaxies. First, it is to generate the spectra of galaxies with different types by advantage of EPS models and various SFR forms, then it is to compute the above mentioned parameters from the generated spectra.
About the descriptions of various EPS models [including the Yunnan, BC03 \citep{bru03}, {\hf STARBURST99} \citep[hereafter {\hf SB99},][]{lei99,lei10,vaz05}, {\hf P\'EGASE} \citep{fio97,fio99} and POPSTAR \citep{mol09}], various SFR forms, the method of building various types of galaxies and the algorithms of obtaining the above mentioned parameters, we have given in Paper I.
Here, we only present the simple descriptions of EPS models, IMFs [$\phi(M)= {\rm d}N / {\rm d}M$] and SFR forms. In Table~\ref{Tab:mod-des}, we present the name, the corresponding IMFs, the lower and upper mass limits ($M_{\rm l}$ and $M_{\rm u}$) and metallicities for each set of models in the second, third, fourth and fifth columns, respectively.

\subsection{EPS models}
As said above, the detailed description of various EPS models has been presented in Paper I, we refer the interested reader to part 2 for them.
In Paper I we only use solar-metallicity EPS models and present the above SFR calibrations at solar metallicity. In this paper, we use the EPS models at several metallicities (see the fifth column of Table~\ref{Tab:mod-des}, which gives the heavy-element abundance by mass $Z$), discuss the effects of binary interactions, metallicity, EPS models and IMF on these SFR calibrations.

Moreover, for the {\hf P\'EGASE} EPS models, we do not use the default (i.e. consistent) evolution process of stellar metallicity, but present the results at individual metallicities.

\subsection{IMFs}
\label{Sect:imf}
\begin{itemize}
\item In the Yunnan models, the IMF of \citet[][hereafter MS79]{mil79} is used, its form is as follows:
\begin{equation}
\phi(M)_{_{\rm MS79}} \propto \Biggl\{ \matrix{
          M^{-1.4}, & 0.10 \le M \le 1.00, \cr
          M^{-2.5}, & 1.00 \le M \le 10.0, \cr
          M^{-3.3}, & 10.0 \le M \le 100, \cr
          }
\label{eq.imfms79}
\end{equation}
where $M$ is the stellar mass in units of M$_{\rm \odot}$.

\item In the BC03, {\hf SB99, P\'EGASE} and POPSTAR models, the \citet[][hereafter S55]{sal55} IMF is used and its form is as follows: $\phi(M)_{_{\rm S55}} = M^{-\alpha}$, $\alpha=2.35$ and the lower and upper mass limits are 0.1 (except for the POPSTAR models) and 100.\,$\rm M_{\odot}$. In the POPSTAR models, the lower mass limit of the S55 IMF is 0.15\,M$_\odot$, which is different from that of the other EPS models, we call S55' IMF in Table~\ref{Tab:mod-des}.

\item In the BC03 models, the used \citet[][hereafter Cha03]{cha03} IMF is as follows:
\begin{equation}
\phi(M)_{_{\rm Cha03}} = \Bigl\{ \matrix {
          {\rm C_1} M^{-1} {\rm exp}^ {[{-({\rm log}M-{\rm logM_c})^2 \over 2\sigma^2}]}, & M \le 1.0, \cr
          {\rm C_2} M^{-2.3} \hfill,  & M > 1.0, \cr
          }
\label{eq.imfcha03}
\end{equation}
where ${\rm M_c}=0.08$\,M$_\odot$, $\sigma=0.69$ and $M$ is the stellar mass in units of M$_\odot$. The lower and upper mass limits are 0.1 and 100.$\rm M_\odot$.

\item The IMF of \citet[][hereafter K93]{kro93}, which is used in the {\hf SB99 and P\'EGASE} models, is as follows:
\begin{equation}
\phi(M)_{_{\rm K93}} = \Biggl\{ \matrix {
          {\rm C_1} M^{-1.3}, \ \ 0.10 \le M \le 0.50, \cr
          {\rm C_2} M^{-2.2}, \ \ 0.50 \le M \le 1.00, \cr
          {\rm C_3} M^{-2.7}, \ \ 1.00 \le M \le 100, \cr
          }
\label{eq.imfk93}
\end{equation}
where ${\rm C_1}=0.035$, ${\rm C_2}=0.019$, ${\rm C_3}=0.019$ and $M$ is the stellar mass in units of M$_{\odot}$. Because all coefficients in equation (\ref{eq.imfk93}) are set to 1 for the {\hf SB99} models in this study and also are 1 in the {\hf P\'{E}GASE} models, we call K93' IMF in Table~\ref{Tab:mod-des}.

\item The IMF of \citet[][ hereafter K01]{kro01}, which is used in the POPSTAR models, is as follows:
\begin{equation}
\phi(M)_{_{\rm K01}} = \Biggl\{ \matrix {
          {\rm C_1} M^{-0.30}, & 0.01 \le M \le 0.08, \cr
          {\rm C_2} M^{-1.30}, & 0.08 \le M \le 0.50, \cr
          {\rm C_3} M^{-2.30}, & 0.50 \le M \le 100, \cr
          }
\label{eq.imfk01}
\end{equation}
where $M$ is the stellar mass in units of M$_\odot$. In the POPSTAR models, $M_{\rm l}$ = 0.15\,M$_{\rm \odot}$.
\end{itemize}

\subsection{SFR forms}
\label{Sect:csp}
Various SFR forms are used to transform SP to galaxies with different types. We use a $\delta$-form SFR, six exponentially decreasing SFRs with characteristic time decays $\tau =1, 2, 3, 5, 15$ and 30\,Gyr and a constant-form SFR to build burst, E, S0, Sa-Sd and Irr types of galaxies, respectively.
The exponentially decreasing SFR is given by
\begin{equation}
\psi(t) = [1 + \epsilon M_{\rm PG} (t)] \tau ^{-1} {\rm exp}(-t/\tau),
\label{eq.sfr}
\end{equation}
where $\tau$ is the e-folding time-scale, $M_{\rm PG} (t)$ = $[1-{\rm exp}(-t/\tau)] - M_{\rm stars} - M_{\rm remnants}$ is the mass of gas that has been processed into stars and then returned to the ISM at $t$, $M_{\rm stars}$ and $M_{\rm remnants}$ are the masses of stars and remnants at $t$, and $\varepsilon$ denotes the fraction of $M_{\rm {PG}} (t)$ that can be recycled into new star formation.
In this work, $\varepsilon = 0.$, i.e. the gas could not be recycled into new star formation.

\subsection{the dependence of calculations on metallicity}
\label{Sect:csp-z}
In the transformation between the number of ionizing photons $Q$(H) and $L_{\rm H\alpha}$,
\begin{equation}
L_{\rm H\alpha} =Q(\rm H) {\alpha \over \beta} \, {j_{\rm B} \over \alpha_{\rm B}}, 
\label{Eq:lha}
\end{equation}
we assume case B recombination at election temperature $T_e = 10\,000$K and number density $n_e = 100\,$cm$^{-3}$ at all metallicities, i.e. do not consider the effect of metallicity.

In fact, using the data of Table 1 from \citet{zhao10}, we find that the electron temperature $T_e$ (in the range of 10\,000$-$20\,000\,K) decreases linearly with the Oxygen abundance 12+log(O/H) (in the range of 7.4$-$8.6) for blue compact dwarf galaxies, and the number density $n_e$ is generally lower than 100\,cm$^{-3}$ except for at low Oxygen abundance [12+log(O/H)=7.41].
However, from Table 5 of \citet{fer80}, we see that the Balmer decrement $\alpha / \beta$ only increases by 0.033 times (from 2.69 to 2.78) when $T_e$ changes from 10\,000 to 20\,000K. 
Moreover, from equations (6) and (8) of \citet{fer80}, we see that recombination coefficient $\alpha_{\rm B}$ and emission coefficient $j_{\rm B}$ are proportional to $T_e^{-0.77}$ and $T_e^{-0.833}$ when $T_e \le 2.6 \times 10^4$K,  respectively, $j_{\rm B} / \alpha_{\rm B}$ decreases by 0.043 times when $T_e$ changes from 10\,000 to 20\,000K. 
Hence, the conversion factor ($\alpha \over \beta$\,$j_{\rm B} \over \alpha_{\rm B}$) between $Q$(H) and $L_{\rm H\alpha}$ is almost invariable when $T_e$ changes from 10\,000 to 20\,000K (decreases by 0.014 times), it is reasonable to assume that $\alpha \over \beta$\,$j_{\rm B} \over \alpha_{\rm B}$ is independent of metallicity.

We also do not consider the effect of metallicity on the ratio of $L_{\rm [OII]}$ to $L_{\rm H\alpha}$ (=0.23 in Paper I, which is the conclusion made by \citealt{hop03}) for the sake of its uncertainty.
In the work of \citet{kew03}, they made the conclusion that the ratio of $L_{\rm [OII]}$ to $L_{\rm H\alpha}$ is dependent of metallicity.

\section{Some results about binary evolutions}
\begin{figure*}
\leftline{
\includegraphics[bb=108 34 532 715,clip,angle=270,scale=.300]{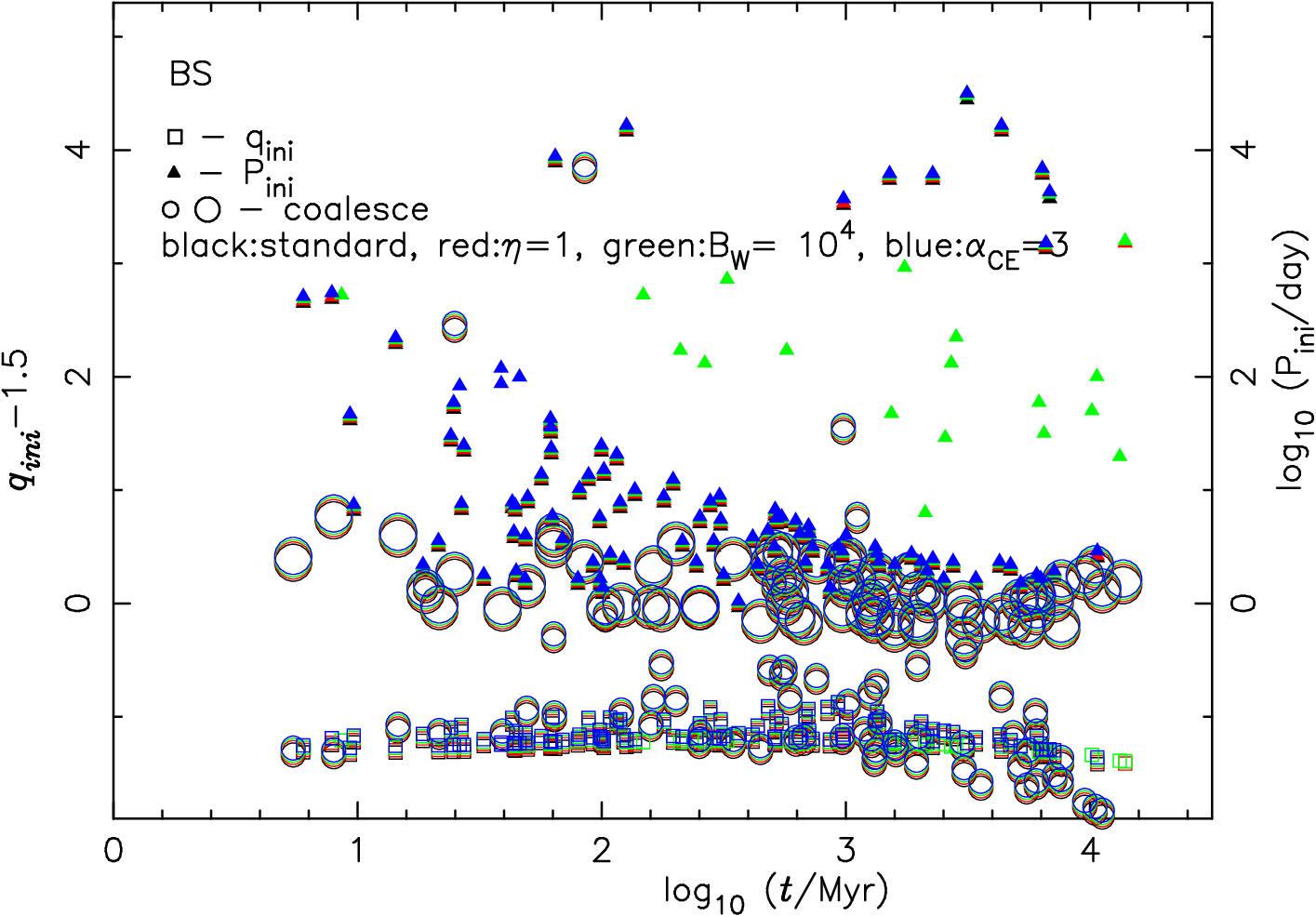}
\includegraphics[bb=108 34 532 715,clip,angle=270,scale=.300]{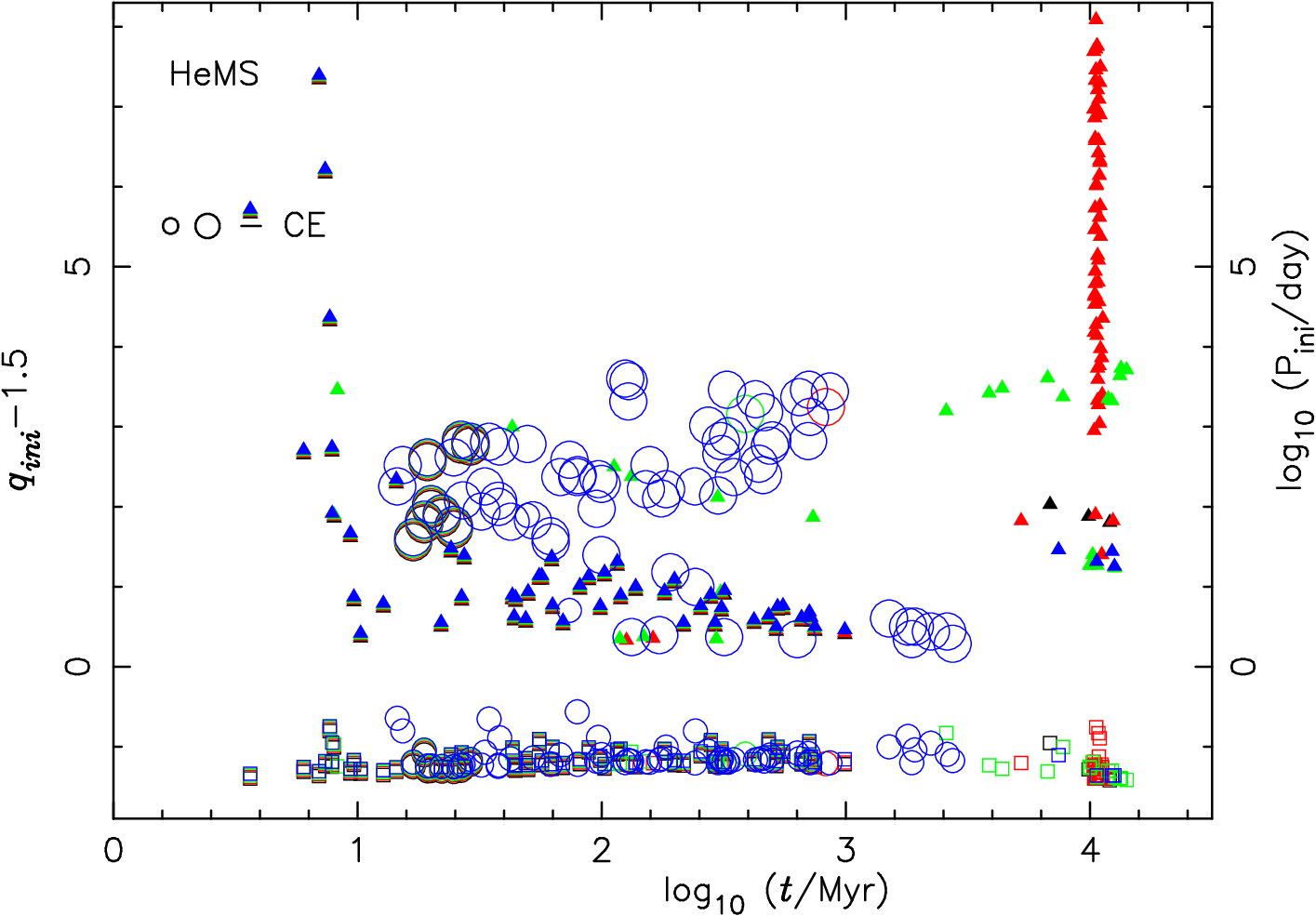} 
}
\leftline{
\includegraphics[bb=108 34 532 715,clip,angle=270,scale=.300]{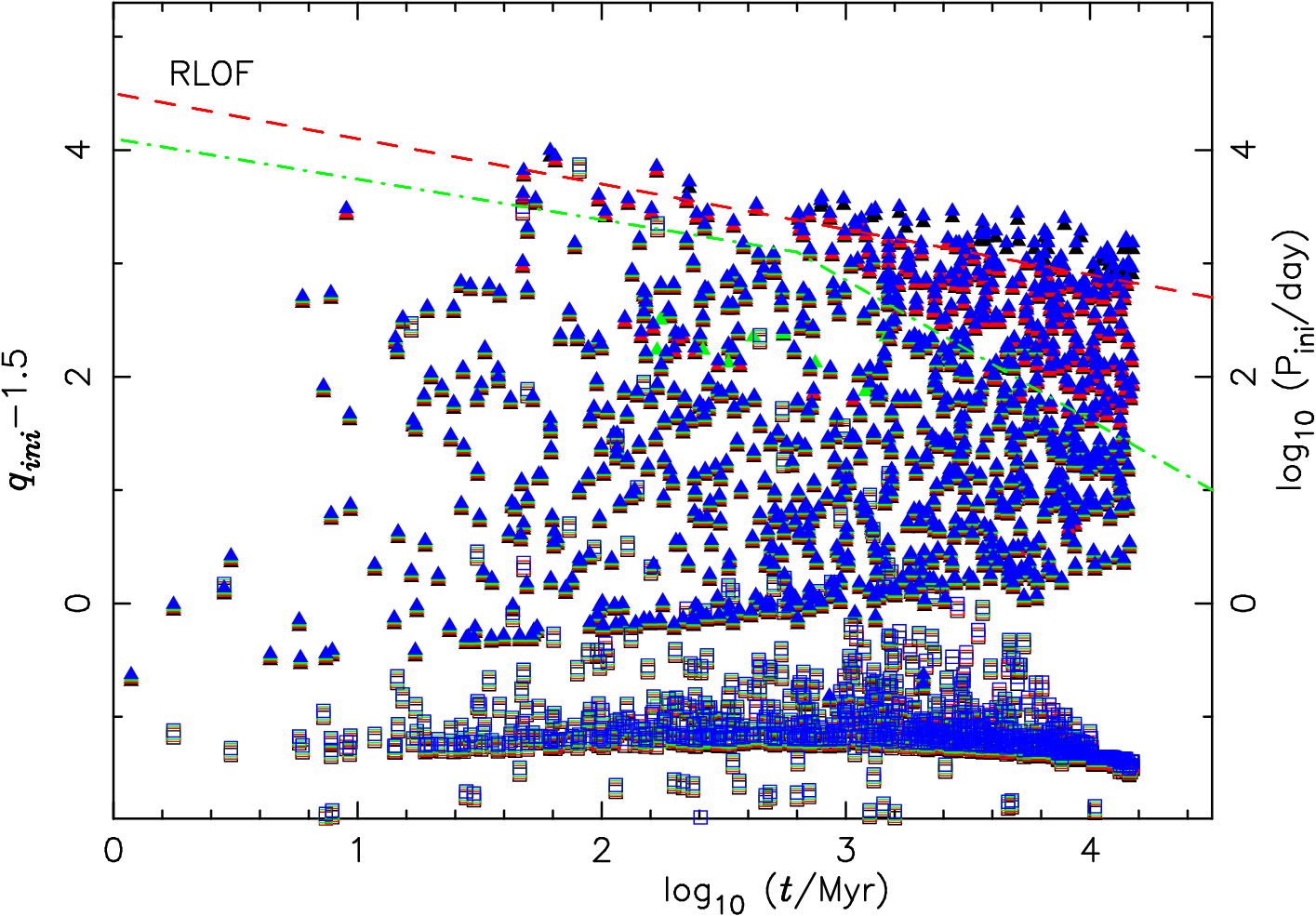}
\includegraphics[bb=108 34 532 715,       angle=270,scale=.300]{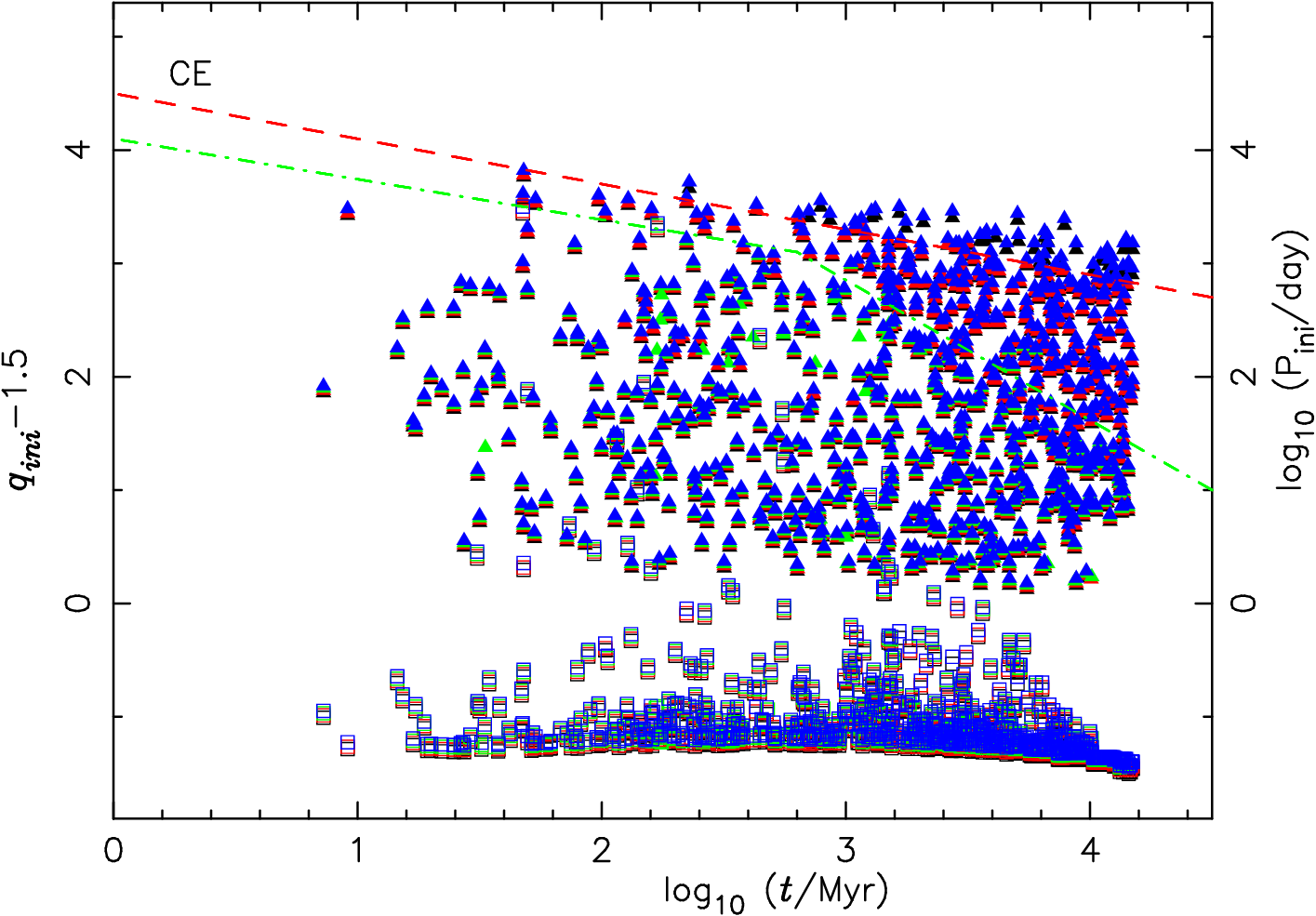} 
}
\leftline{
\includegraphics[bb=108 34 572 715,       angle=270,scale=.300]{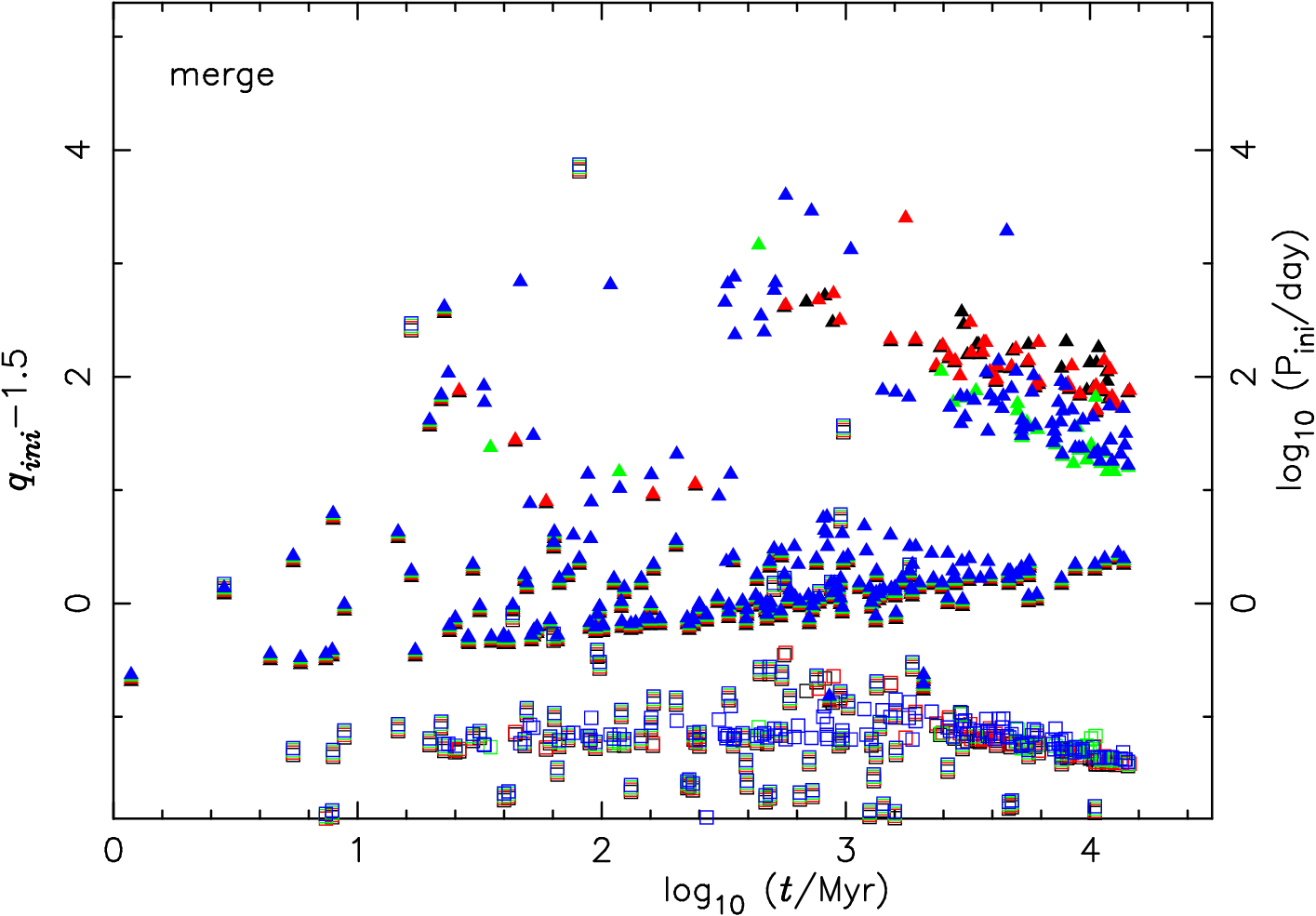} 
}
\caption{The initial mass ratio (left y-axis, $q_{\rm ini}-$1.5, open rectangles) and period [right y-axis, log$_{10}$($P_{\rm ini}$/day), solid triangles] of binaries as a function of the onset-time of BSs (excluding the systems composed of BS and WD), HeMS stars, RLOF, CE and merge processes (top-left, top-right, intermediate-left, intermediate-right and bottom-left panels).  In each panel, the black, red, green and blue symbols are for the cases of standard, $\eta=1.0$, B$_{\rm W}=10000$ and $\alpha=3$. Moreover, the small and large open circles in the top-left and top-right panels represent mass ratio and period of BSs via coalesce and HeMS stars via CE ejection processes, respectively. 
The meanings of red dashed and green dot-dashed lines in the middle panels are explained in Section 3.3.
For the sake of clarity, the symbols are moved upwards in different cases. These results are obtained at solar metallicity.}
\label{Fig:rst-bi}
\end{figure*}

\begin{table*}
\centering
\caption{The percentages of BSs ($f_{\rm BS}$), HeMS stars ($f_{\rm HeMS}$), systems experiencing RLOF ($f_{\rm RLOF}$), CE ($f_{\rm CE}$) and merge ($f_{\rm merge}$) processes during the past 13.7\,Gyr. 
Also the per cent of BSs via coalesce channel ($f_{\rm BS, cl}$) and ${f_{\rm BS,cl} \over f_{\rm BS}}$ are presented in the bracket of the second column, 
the per cent of HeMS stars via CE ejection channel ($f_{\rm HeMS, CE}$) and ${f_{\rm HeMS,CE} \over f_{\rm HeMS}}$ are presented in the bracket of the third column.
In the second column the last value is for the systems composed of BS and non-WD stars ($f_{\rm BS, non-WD}$).
{\bf Top part} is for the standard models and metallicity $Z$=0.0001, 0.0003, 0.001, 0.004, 0.01, 0.02 and 0.03 (from top to bottom). {\bf Bottom part} is at solar metallicity and in the cases of standard, $\eta=1.0$, B$_{\rm W}=10^4$ and $\alpha_{\rm CE}$=3.0 (from top to bottom).
}
\begin{tabular}{lcc ccc}
\hline
\hline
\multicolumn{6}{c}{standard models} \\

  $Z$ & $f_{\rm BS} (f_{\rm BS,cl}, {f_{\rm BS,cl} \over f_{\rm BS}} ) ,f_{\rm BS, non-WD}$  & $f_{\rm HeMS} (f_{\rm HeMS,CE}, {f_{\rm HeMS,CE} \over f_{\rm HeMS}}) $ & $f_{\rm RLOF}$ & $f_{\rm CE}$ & $f_{\rm merge}$\\
              & \%(\%,\%),\% & \%(\%,\%),\% & (\%) & (\%) & (\%)   \\
\hline
$0.0001$    &   3.58(.42,  12), 2.43	   &   1.33(.11,  8.)	   &   10.59   &   9.13	   &   2.03 \\
$0.0003$    &   3.46(.52,  15), 2.35	   &   1.36(.16,  12)	   &   10.70   &   9.23	   &   2.05 \\
$0.001$    &   3.66(.53,  15), 2.20	   &   1.24(.21,  17)	   &   10.74   &   9.28	   &   1.97 \\
$0.004$    &   2.94(.62,  21), 1.84	   &   0.95(.15,  16)	   &   10.42   &   8.81	   &   2.07 \\
$0.01  $    &   2.69(.65,  24), 1.78	   &   0.74(.10,  14)	   &   9.83	   &   8.14	   &   2.10 \\
$0.02  $  &   2.50(.68,  27), 1.64	   &   0.62(.08,  13)	   &   9.48	   &   7.71	   &   2.11 \\
$0.03  $  &   2.49(.65,  26), 1.52	   &   0.61(.08,  13)	   &   9.28	   &   7.50	   &   2.15 \\

\hline
\multicolumn{6}{c}{$Z$=0.02} \\
  case & $f_{\rm BS} (f_{\rm BS,cl}, {f_{\rm BS,cl} \over f_{\rm BS}} ) ,f_{\rm BS, non-WD}$  & $f_{\rm HeMS} (f_{\rm HeMS,CE}, {f_{\rm HeMS,CE} \over f_{\rm HeMS}}) $ & $f_{\rm RLOF}$ & $f_{\rm CE}$ & $f_{\rm merge}$\\
\hline
standard	                     &   2.50(.68,  27), 1.64	   &   0.62(.08,  13)	   &   9.48	   &   7.71	   &   2.11 \\
$\eta$=1.0	            &   2.52(.68,  27), 1.63	   &   1.20(.09,   8.)	   &   9.06	   &   7.32	   &   2.09 \\
B$_{\rm W}$=$10^4$ &   3.22(.68,  21), 1.78	   &   0.83(.09,  11)	   &   6.89	   &   4.87	   &   1.93 \\
$\alpha_{\rm CE}$=3 &   2.59(.68,  27), 1.68	   &   1.25(.70,  56)	   &   9.48	   &   7.71	   &   2.68 \\
 \hline
\end{tabular}
\label{Tab:rst-bi}
\end{table*}

Before discussing the effects of metallicity, binary interactions, EPS models and IMF on the SFR calibrations, we first give some descriptions of formation channels for some classes of objects and results concerning binary evolutions. In our works, the binary star evolution (BSE) code of  \citet{hur02} has been used.
Using the same set of input parameters and physics as in our works,  \citet{hur02} used the BSE code  to compare the model results with the observations for many objects [including blue stragglers (BSs), Algol, CVs, X-ray and so on] and found that they match well.

In the populations we constructed, if the component stars in a binary system are close enough, they would interact with each other and experience processes such as mass transfer, mass accretion, common-envelope (CE) evolution, collision, supernova kick, tidal evolution, angular momentum loss and so on. 
As a consequence, some of binaries would evolve to/through the systems comprising high-temperature and high-luminosity star [for example, BSs and helium main-sequence (HeMS) stars] or these kinds of single stars. These systems would significantly alter the spectra of SPs. 
Among the above-mentioned processes, Roche lobe overflow (RLOF), CE and merge processes are the most important ones to alter the evolution sequences of stars as expected from single star evolution.
In the following, we will describe the formation channels of BSs and HeMS stars (including in binary and single systems), the percentages of BSs, HeMS stars and those systems experiencing RLOF, CE and merger processes during the past 13.7\,Gyr, and the dependences of these results on some parameters (including metallicity, stellar wind and CE ejection coefficient $\alpha_{\rm CE}$=$\Delta E_{\rm bind}/\Delta E_{\rm orb}$, $\Delta E_{\rm bind}$ and $\Delta E_{\rm orb}$ are the energy added to the binding energy of the envelope and the change in the orbital energy of the binary between the initial and final states of the spiraling-in process) in our models. 
In this section, the results and conclusions are obtained by using SPs composed of $10^4$ binary systems.
However, the results and conclusion in Section 4 are based on the SPs comprising 2.5$\times$$10^7$ binary systems.

In our models, Reimers mass loss coefficient $\eta$ is taken as 0.3, tidally enhanced mass loss coefficient B$_{\rm W}$ is set constant at 0 and CE ejection coefficient $\alpha_{\rm CE} = 1.0$.
In order to differ these models from the results when using the other sets of parameters (Section 3.3), we call them the standard models.
Moreover, it is emphasized that the phases from main sequence (MS) to remnant [white dwarf (WD), etc] are included in the BSE code.

\subsection{Formation channels for BSs and HeMS stars}
Using the criterion of $t>t_{\rm (MS, M_0)}$ (i.e. the duration on the MS phase is greater than the MS lifetime for any component star with an initial mass of $M_0$ in a binary system), in our models, BSs can be formed via coalesce (MS+MS) and mass transfer (MS+companion) processes. 
In the case of coalesce, two sub-channels are included: $RLOF \rightarrow contact \rightarrow merge$ and $RLOF \rightarrow merge$. 
By mass transfer channel, MS star can stably accrete companion's mass via RLOF or accrete companion's wind. The latter case (wind) produces the relatively low luminosity BSs because of a small amount of accreted material. This has been confirmed by \citet{pol94} and \citet{hur05}.
In Fig.~\ref{Fig:rst-bi}, we present the initial mass ratio ($q_{\rm ini}-$1.5) and period [log$_{10}$($P_{\rm ini}$/day)] of binaries as a function of the onset-time, at this time the binary system begin to experience BS, HeMS, RLOF, CE and merge phases/processes, in the cases of standard, $\eta=1$, B$_{\rm W}=10^4$ and $\alpha_{\rm CE}=3$.
From the top-left panel, we see that BSs via coalesce channel (open circles) mainly origin from short period binaries and the number of BSs via coalesce channel is less ($\sim12-$27 per cent for the standard models from $Z$=0.0001 to 0.03) than that via mass transfer channel.

In our models, HeMS stars are mainly produced by mass loss [Hertzsprung gap$\rightarrow$HeMS, the first giant branch (GB)$\rightarrow$HeMS, core helium burning$\rightarrow$HeMS] and CE ejection processes. In the top-right panel, the HeMS stars via CE ejection process are represented by open circles. In our standard models,  only $\sim8-$17 per cent of HeMS stars are formed via CE ejection channel.

\subsection{Percentages of BSs, HeMS stars, binaries experiencing RLOF, CE and merge processes and the dependence on $Z$}
In the top part of Table~\ref{Tab:rst-bi}, we give the percentages of BSs ($f_{\rm BS}$), HeMS stars ($f_{\rm HeMS}$), the binaries experiencing RLOF ($f_{\rm RLOF}$), CE ($f_{\rm CE}$) and merge ($f_{\rm merge}$) processes during the past 13.7\,Gyr for the standard models at metallicity $Z$=0.0001, 0.0003, 0.001, 0.004, 0.01, 0.02 and 0.03 (the first column, from top to bottom).  
Also we give the per cent of BSs via coalesce channel ($f_{\rm BS, cl}$) and ${f_{\rm BS,cl} \over f_{\rm BS}}$ in the bracket of the second column, 
the per cent of HeMS stars via CE ejection channel ($f_{\rm HeMS, CE}$) and ${f_{\rm HeMS,CE} \over f_{\rm HeMS}}$ in the bracket of the third column.
At last, in the second column, the per cent of the systems composed of BS and non-WD stars ($f_{\rm BS, non-WD}$) is also presented after $f_{\rm BS}$.
The percentage equals to ${N_{\rm case} \over N_{\rm ini, tot}} \times 100, N_{\rm case}$ is the total number of  binaries experiencing the corresponding case during the past 13.7\,Gyr and $N_{\rm ini,tot}$  is the total number of binary systems at the initial condition (i.e. $=10^4$).

From the top part of Table~\ref{Tab:rst-bi}, we see that $f_{\rm BS}$, $f_{\rm HeMS}$,  $f_{\rm RLOF}$ and $f_{\rm CE}$ decrease with metallicity, but $f_{\rm merge}$ almost does not vary ($\sim$ 2.0\%, shows a slight increase).
Furthermore, from the values in the brackets, we see that the number of BSs via coalesce channel $f_{\rm BS, cl}$ and ${f_{\rm BS, cl} \over f_{\rm BS}}$ increase, but  that via mass transfer channel ($\simeq f_{\rm BS} - f_{\rm BS, cl}$) decreases with metallicity.
For HeMS stars, the number via CE ejection channel $f_{\rm HeMS, CE}$ and  ${f_{\rm HeMS,CE} \over f_{\rm HeMS}}$ increase at low- then decrease at high-metallicity ranges, but that via mass loss channel ($\simeq f_{\rm HeMS} - f_{\rm HeMS, CE}$) decreases when increasing $Z$.

\subsection {The dependences on stellar wind and CE ejection coefficient $\alpha_{\rm CE}$}
We know that stellar wind is a very important parameter during stellar evolutions, especially for massive stars. 
In the BSE code, several descriptions of mass loss are used, including 
(i) that of \citet{rei75} for intermediate- and low-mass stars on the GB and beyond ($M_{\rm R} = \eta \  4 \times 10^{-13} {LR \over M}$, $\eta$ is Reimers mass loss coefficient, $L, R$ and $M$ are luminosity, radius and mass in solar units, respectively); 
(ii) that of \citet{nie90} for massive stars;
 (iii) that of \citet{vas93} for pulsation-driven wind on the asymptotic giant branch (AGB); 
 (iv) that of \citet{tou88} for tidally enhanced mass loss,
 \begin{equation}
 M = M_{\rm R} \Bigl[1+{\rm B_W} \, {\rm min}({1 \over 2}, { R\over R_{\rm L}})^6 \Bigl],
 \label{Eq:bw} 
 \end{equation}
 in which $M_{\rm R}$ is Reimers mass loss, B$_{\rm W}$ and $R_{\rm L}$ are the coefficient and RLOF radius in solar units; 
 (v) that of \citet{ham98} for Wolf-Rayet-like mass loss of stars with small H-envelope mass and 
 (vi) that of \citet{hum94} for luminous-blue-variable-like mass loss beyond the Humphreys-Davidson limit.

Moreover, CE evolution is one of the most important and complex but also one of the least understood phases of binary evolution.
In the BSE code, a widely used and relatively simple criterion of CE evolution is used, i.e. the CE is ejected when $\Delta E_{\rm orb} \ \alpha_{\rm CE}$ exceeds $\Delta E_{\rm bind}$.

In the following, we will discuss the effects of stellar wind (including Reimers and tidally enhanced mass losses) and CE ejection coefficient on the above results. These results are obtained at solar metallicity.
In the bottom part of Table~\ref{Tab:rst-bi}, we give $f_{\rm BS}$ ($f_{\rm BS,cl}$), $f_{\rm HeMS}$ ($f_{\rm HeMS, CE}$), $f_{\rm RLOF}$, $f_{\rm CE}$ and $f_{\rm merge}$  in the cases of standard, $\eta=$ 1.0, B$_{\rm W}=10^4$ and $\alpha_{\rm CE}=3.0$ at solar metallicity (from the first to the last lines).  
The conclusions and analyses are as follows.

\subsubsection{Reimers mass loss coefficient $\eta$}
From the bottom part of Table~\ref{Tab:rst-bi}, we see that when $\eta$ is from 0.3 to 1.0, $f_{\rm BS}$ and $f_{\rm BS, cl}$ almost do not vary,  $f_{\rm HeMS}$ increases significantly but $f_{\rm HeMS, CE}$ almost does not vary, $f_{\rm RLOF}$ and $f_{\rm CE}$  decrease and $f_{\rm merge}$ almost does not vary.

From Fig.~\ref{Fig:rst-bi}, we can see clearly the variations in $f_{\rm BS}$, $f_{\rm HeMS}$, $f_{\rm RLOF}$, $f_{\rm CE}$ and $f_{\rm merge}$ caused by the increase of $\eta$.
{\bf (i)} From the top-right panel of Fig.~\ref{Fig:rst-bi}, it can be seen that the increase of $\eta$ raises the number of HeMS stars at an age of $t \sim 10^9$\,yr, these HeMS stars origin from long period binaries [log$_{10}$($P_{\rm ini}$/day)$\ga$3.] and are formed from mass loss process (see the second paragraph of Section 3.1, $\simeq f_{\rm HeMS} - f_{\rm HeMS, CE}$). The number of HeMS stars via CE ejection process ($f_{\rm HeMS, CE}$, represented by open circles in the top-right panel of Fig.~\ref{Fig:rst-bi}) almost does not vary, and these HeMS stars are formed from binaries with log$_{10}$($P_{\rm ini}$/day)$\sim$2 and at early ages ($t \sim 10^7$\,yr) in the standard and $\eta=1.0$ cases.
{\bf (ii)} From the intermediate panels of Fig.~\ref{Fig:rst-bi}, we see that the increase of $\eta$ mainly decreases $f_{\rm RLOF}$ and $f_{\rm CE}$ in the range of  $t \ga 10^9$\,yr and log$_{10}$($P_{\rm ini}$/day) $\ga$ 3 (the region above the red dashed line). 
Checking the evolutionary stage, we find that the binaries located above the red dashed line in the standard models have evolved to the systems with post thermal pulsing AGB (TP-AGB) component at the onset-time for RLOF process and those with HeWD component for CE process.

The analyses are as follows. Reimers mass loss and its coefficient $\eta$ is valid for intermediate- and low-mass stars on the RGB and beyond. Therefore, the variations in the $f_{\rm BS}$, $f_{\rm HeMS}$, $f_{\rm RLOF}$, $f_{\rm CE}$ and $f_{\rm merge}$ caused by the increase of $\eta$ are at intermediate and large ages. 
{\bf (i)} The increase in $\eta$ raises the envelope loss efficiency of component star (for examples, on the RGB, HeMS phases) in a binary system, thus raises the number of HeMS stars via mass loss channel ($\simeq f_{\rm HeMS} - f_{\rm HeMS, CE}$) and $f_{\rm HeMS}$, while almost does not vary $f_{\rm HeMS, CE}$. 
{\bf (ii)} The BSs are produced mainly via MS-MS merge and mass transfer processes, the accretion from companion's wind can not increase significantly the mass of MS star, so the increase of $\eta$ does not raise the number of BSs via mass transfer process ($\simeq f_{\rm BS} -f_{\rm BS, cl}$, see the top-left panel and the first paragraph of Section 3.1) and $f_{\rm BS}$.
{\bf (iii)} The RLOF process happens when $R \ge R_{\rm L}$ and CE process happens when $R \ge R_{\rm L}$ and $q  > q_{\rm crit}$ ($q_{\rm crit}$ is the critical mass ratio). When $\eta$ increases, the envelop of component star on the post-TPAGB phase is easily driven away, this lowers the possibility of $R \ge R_{\rm L}$. Therefore, the increase of $\eta$ would lower $f_{\rm RLOF}$ and $f_{\rm CE}$ (the binaries with HeWD) at long period ranges.
{\bf (iv)} The merge process mainly happens at relatively short period ranges (seen the bottom panel or the top-left panel of Fig.~\ref{Fig:rst-bi}), the variation in $\eta$ mainly affects the results within long period ranges, so it almost does not affect $f_{\rm merge}$.

\subsubsection{Tidally enhanced mass loss  coefficient B$_{\rm W}$}
From the bottom part of Table~\ref{Tab:rst-bi}, we see that when B$_{\rm W}$ is from 0. to 10$^4$, both $f_{\rm BS}$ (more significantly) and $f_{\rm HeMS}$ increase while $f_{\rm BS, cl}$ and $f_{\rm HeMS, CE}$ almost do not change,  both $f_{\rm RLOF}$ and $f_{\rm CE}$ decrease, and $f_{\rm merge}$ slightly decreases.
Comparing with the variations caused by increasing $\eta$, we find that the variation trends of $f_{\rm BS, cl}$, $f_{\rm HeMS}$, $f_{\rm HeMS, CE}$, $f_{\rm RLOF}$, $f_{\rm CE}$ (except $f_{\rm BS}$) caused by the inclusion of tidally enhance mass loss are similar to those caused by increasing $\eta$, however, the variation in $f_{\rm HeMS}$ is relatively small and those in  $f_{\rm RLOF}$ and $f_{\rm CE}$ are relatively large.
Moreover, excluding the systems with BS and WD stars (see the last number in the second column of Table~\ref{Tab:rst-bi}), $f_{\rm BS}$ does not increase so much.

Also, from Fig.~\ref{Fig:rst-bi}, we can see clearly the variations caused by the inclusion of tidally enhance mass loss.
{\bf (i)} From the top panels of Fig.~\ref{Fig:rst-bi}, we see that the inclusion of tidally enhanced mass loss raises the number of BSs via mass transfer process ($\simeq f_{\rm BS}-f_{\rm BS,cl}$) and HeMS stars via mass loss process ($\simeq f_{\rm HeMS} - f_{\rm HeMS,CE}$). These BSs are evolved from binaries with intermediate period [$1. \la {\rm log_{10}}(P_{\rm ini}/{\rm day}) \la 3. $] and formed at ages $t \ga 10^8$\,yr, and these HeMS stars are originated from binaries with ${\rm log_{10}}(P_{\rm ini}/{\rm day}) \sim 3. 5$ and formed at $t$ $\sim$ $10^9$\,yr.
{\bf (ii)} From the intermediate panels of Fig.~\ref{Fig:rst-bi}, it can be seen that the inclusion of tidally enhanced mass loss decreases the number of binaries experiencing RLOF and CE processes at large ages, these binaries origin from systems with long period (the region above the green dot-dashed line). 
Check the evolutionary stage, the binaries above the green dot-dashed line have evolved to the systems with GB/post-TP-AGB component for RLOF process and those with GB/HeWD component for CE process.
Comparing with the results of $\eta=1.0$, the reduced period region is larger.
{\bf (iii)} The merge process happens at short period ranges when considering tidally enhanced mass loss.

The reasons are as follows. Tidally enhanced mass loss can be considered as the enhanced version of Reimers mass loss (see equation~\ref{Eq:bw}). It is valid on the RGB and beyond, so the variations caused by the inclusion of tidally enhance mass loss also are at intermediate and large ages. It can raise the number of BSs via mass transfer channel ($\simeq f_{\rm BS} - f_{\rm BS,cl}$) and $f_{\rm BS}$ (but almost has no effect when increasing $\eta$),
and lead to larger variations in the $f_{\rm RLOF}$ and $f_{\rm CE}$.
Why $f_{\rm HeMS}$ is less than that caused by increasing $\eta$? This is because that tidal wind becomes to be significant only when $R$ is comparable to $R_{\rm L}$ (see equation~\ref{Eq:bw}), otherwise it equals to Reimers mass loss with $\eta=0.3$.

\subsubsection{CE ejection coefficient $\alpha_{\rm CE}$}
From the bottom part of Table~\ref{Tab:rst-bi}, we see that the increase of $\alpha_{\rm CE}$ raises $f_{\rm HeMS}$, $f_{\rm HeMS, CE}$ and $f_{\rm merge}$, almost does not change $f_{\rm BS}$, $f_{\rm BS, cl}$, $f_{\rm RLOF}$ and $f_{\rm CE}$.
From the top-right panel of Fig.~\ref{Fig:rst-bi}, we see that it mainly raises the number of HeMS stars via CE process ($f_{\rm HeMS, CE}$), these stars origin from those binary systems with $P_{\rm ini} \sim 10^3$\,day and large $q$ and are formed in the age range of $10^7 \la t{\rm /yr} \la 10^9$.
The reason is that the increase of $\alpha_{\rm CE}$ makes more orbital energy $ E_{\rm orb}$ is transformed to binding energy $E_{\rm bind}$, thus CE is more easily be ejected.


\section{Effects of binary interactions and metallicity on SFR calibrations}
\begin{figure}
\centering
\includegraphics[height=8.0cm,width=6.5cm,angle=270]{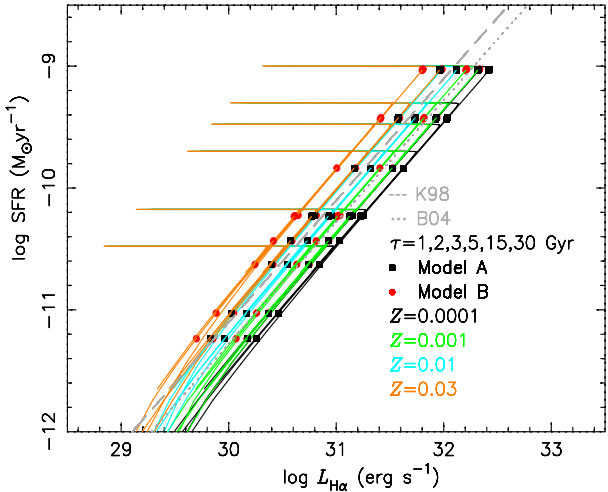}
\caption{Relation between SFR and $L_{\rm H\alpha}$ of E, S0, Sa, Sb, Sc and Sd galaxies (corresponding to $\tau=1,2,3,5,15$ and 30\,Gyr in equation \ref{eq.sfr}, from top to bottom) for Models A (solid rectangles) and B (solid circles) at $Z=0.0001$ (black), 0.001 (green), 0.01 (cyan) and 0.03 (red, from right to left). The ages of galaxies are in the range from 0.1\,Myr to 15\,Gyr. Also shown are the results of K98 (grey dashed line) and B04 (grey dotted line).}
\label{Fig:sfr-lha}
\end{figure}

\begin{figure*}
\includegraphics[angle=270,scale=.410]{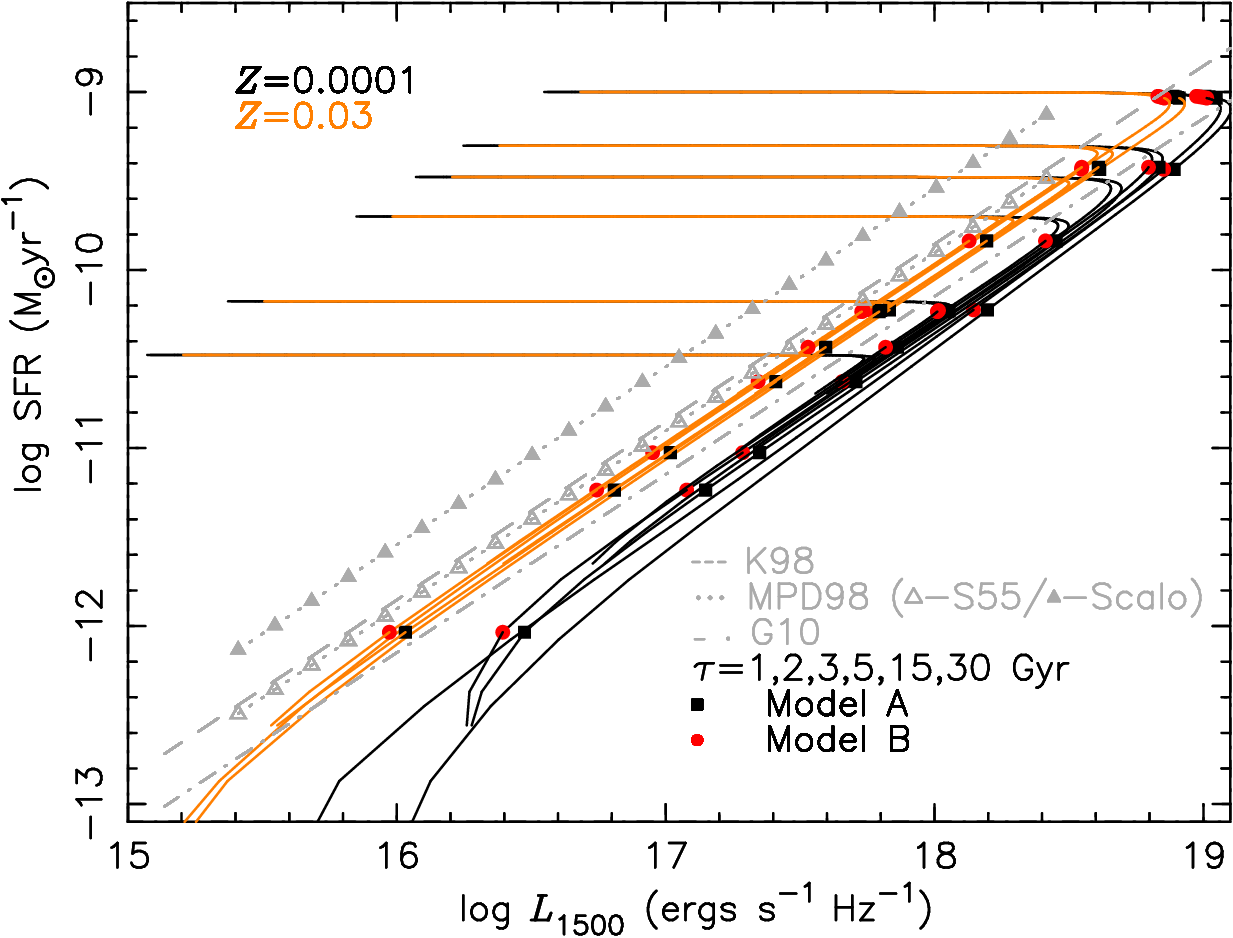}
\includegraphics[angle=270,scale=.410]{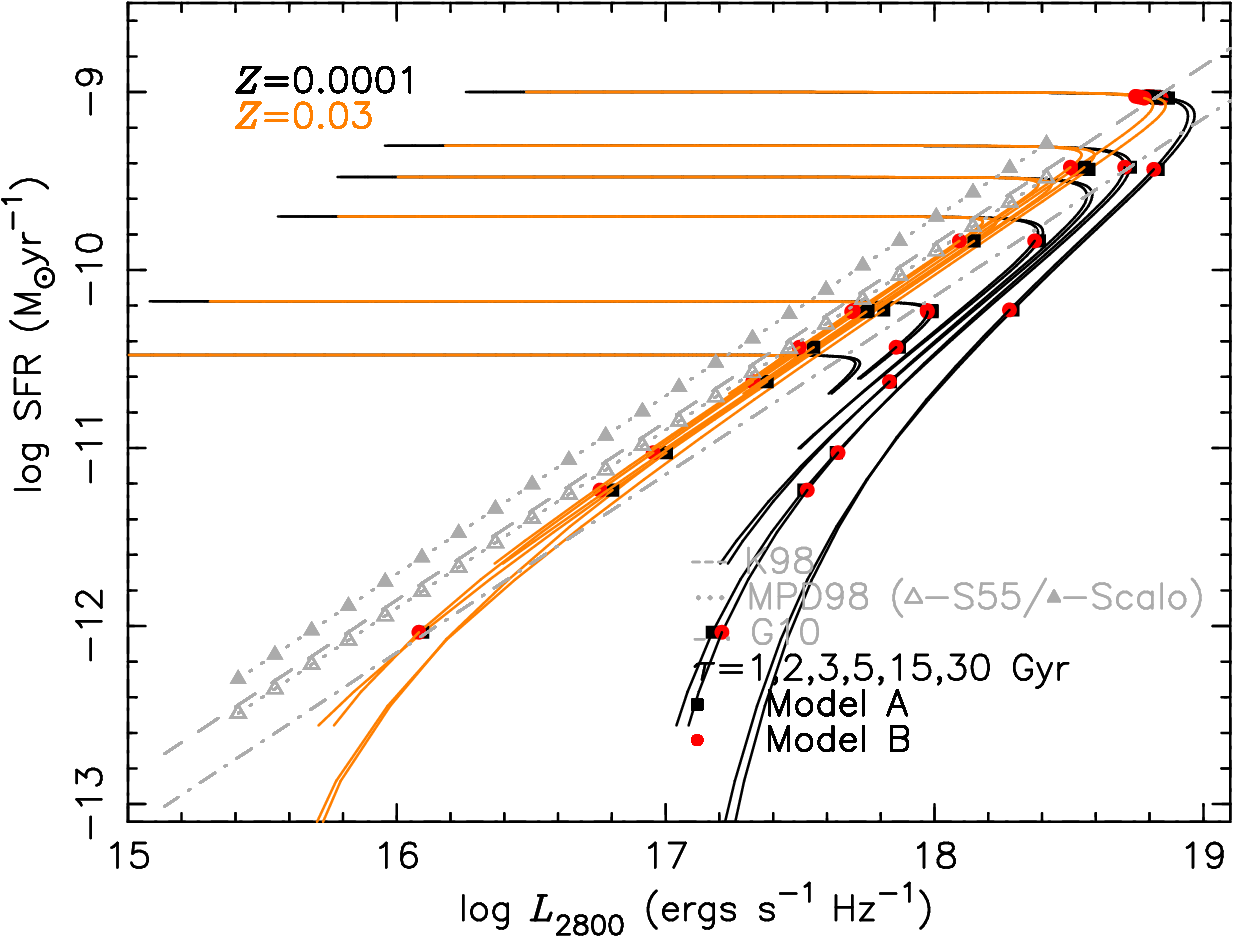}
\caption{Relations between SFR and $L_{i,{\rm UV}}$ of E, S0, Sa, Sb, Sc and Sd galaxies (corresponding to $\tau=1,2,3,5,15$ and 30\,Gyr in equation \ref{eq.sfr}, from top to bottom) for Models A (solid rectangles) and B (solid circles) at $Z=0.0001$ (black) and 0.03 (red, from right to left). Left-hand panel is for $L_{\rm 1500}$ and right-hand panel is for $L_{\rm 2800}$. In each panel, also shown are the results of K98 (grey dashed line), MPD98 (grey dotted line, open and solid triangles are for using the S55 and Scalo IMFs, respectively) and G10 (grey dot-dashed line).}
\label{Fig:sfr-luv}
\end{figure*}

\begin{figure}
\centering
\includegraphics[height=8.0cm,width=6.5cm,angle=270]{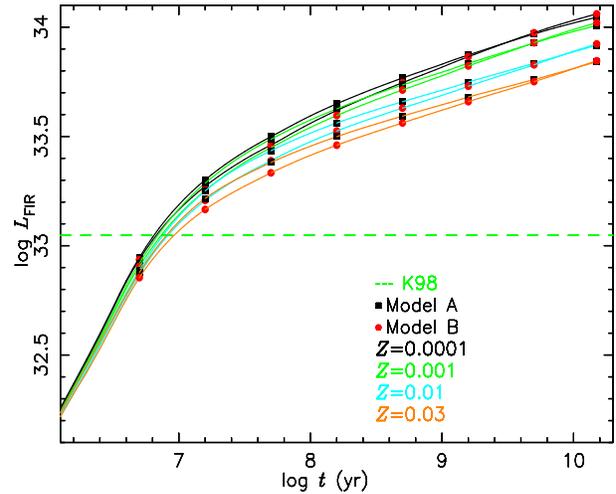}
\caption{The $L_{\rm FIR}$ evolution of Irr galaxies (i.e. models with constant star formation, SFR=1\,M$_{\rm \odot}$) for Models A (solid rectangles) and B (solid circles) at $Z=0.0001$ (black), 0.001 (green), 0.01 (cyan) and 0.03 (red, from top to bottom). Also shown is the result of K98 (green dashed line).}
\label{Fig:sfr-lfir}
\end{figure}

\begin{table*}
\centering
\caption{Conversion coefficient between SFR and $L_{\rm H\alpha}$ (C$_{\rm H\alpha}$, see equation 8) and the rms ($\sigma_{\rm H\alpha}$) for Models A, B, C-S55/Cha03, D-S55/K93', E-S55/K93' and F-S55'/K01 at different metallicities. The top and bottom parts are for Models A-B and C-F, respectively. For each sub-part, the corresponding metallicities are given in the first line (from the second to the last columns).}
\begin{tabular}{llc ccc cc}
\hline
\hline
  Model & \multicolumn{7}{c}{C$_{\rm H\alpha}$, $\sigma_{\rm H\alpha}$}\\
\hline
        & \ \ \ $Z$=0.0001 & $Z$=0.0003 & $Z$=0.001 & $Z$=0.004 & $Z$=0.01 & $Z$=0.02 &$Z$=0.03 \\
      A & $ -41.462,   0.007$  & $ -41.419,   0.007$  & $ -41.361,   0.008$  & $ -41.266,   0.008$  & $ -41.163,   0.008$  & $ -41.056,   0.010$  & $ -41.008,   0.009$ \\
      B & $ -41.366,   0.007$  & $ -41.311,   0.007$  & $ -41.247,   0.008$  & $ -41.129,   0.010$  & $ -41.022,   0.011$  & $ -40.894,   0.014$  & $ -40.849,   0.014$ \\
  & \multicolumn{7}{c}{------------------------------------------------------------------------------------------------------------------------------------------------------------------} \\
        & \ \ \ $Z=$0.0001 & $Z$=0.0004 & $Z$=0.004 & $Z$=0.008 & $Z$=0.02 & $Z$=0.05 & $Z$=0.1 \\
  C-S55 & $ -41.387,   0.006$  & $ -41.282,   0.007$  & $ -41.218,   0.007$  & $ -41.174,   0.007$  & $ -41.078,   0.006$  & $ -40.938,   0.011$  &                     \\
 C-Cha03 & $ -41.608,   0.005$  & $ -41.503,   0.006$  & $ -41.442,   0.006$  & $ -41.398,   0.006$  & $ -41.303,   0.005$  & $ -41.159,   0.010$  &                     \\
   D-S55 &              & $ -41.236,   0.000$  & $ -41.134,   0.000$  & $ -41.066,   0.000$  & $ -40.922,   0.000$  & $ -40.693,   0.000$  &                     \\
   D-K93' &              & $ -41.434,   0.000$  & $ -41.334,   0.000$  & $ -41.267,   0.000$  & $ -41.122,   0.000$  & $ -40.891,   0.000$  &                     \\
   E-S55 & $ -41.384,   0.001$  & $ -41.305,   0.001$  & $ -41.224,   0.001$  & $ -41.160,   0.001$  & $ -41.087,   0.001$  & $ -40.938,   0.001$  & $ -40.950,   0.001$ \\
   E-K93' & $ -41.085,   0.001$  & $ -40.997,   0.001$  & $ -40.899,   0.001$  & $ -40.834,   0.001$  & $ -40.758,   0.001$  & $ -40.617,   0.001$  & $ -40.649,   0.001$ \\
   F-S55' & $ -41.414,   0.285^{\rm UN}$  & $ -41.394,   0.298$  & $ -41.197,   0.213$  & $ -41.136,   0.170$  & $ -41.035,   0.133$  & $ -40.920,   0.090$  &                     \\
   F-K01 & $ -41.595,   0.398^{\rm UN}$  & $ -41.538,   0.399$  & $ -41.332,   0.309$  & $ -41.231,   0.264$  & $ -41.092,   0.189$  & $ -40.986,   0.150$  &                     \\
\hline
\end{tabular}
\label{Tab:c-lha}
\end{table*}

\begin{table*}
\centering
\caption{Differences in the SFR conversion coefficients between Models A and B (i.e. the effect of binary interactions), $\Delta$C$_{\rm case, BI}$, in terms of $L_{\rm H\alpha}$, $L_{\rm 1500}$, $L_{\rm 2800}$,  $L_{\rm [OII]}$ and $L_{\rm FIR}$ at metallicity $Z$=0.0001, 0.0003, 0.001, 0.004, 0.01, 0.02 and 0.03.}
\begin{tabular}{llrrr rrr}
\hline
 $\Delta$C$_{\rm case,BI}$(dex)  & $Z$=0.0001 & 0.0003 & 0.001 & 0.004 & 0.01 & 0.02 & 0.03 \\
\hline
  $\Delta$C$_{\rm H\alpha, BI}$    & $   \ \ 0.096$  & $   0.108$  & $   0.114$  & $   0.138$  & $   0.141$  & $   0.161$  & $   0.160$ \\
  $\Delta$C$_{\rm 1500, BI}$           & $   \ \ 0.045$  & $   0.046$  & $   0.049$  & $   0.056$  & $   0.060$  & $   0.066$  & $   0.066$ \\
  $\Delta$C$_{\rm 2800, BI}$           & $   \ \ 0.013^{\rm UN}$  & $   0.018$  & $   0.027$  & $   0.037$  & $   0.044$  & $   0.051$  & $   \ \ 0.055$ \\
  $\Delta$C$_{\rm [OII], BI}$      & $   \ \ 0.096$  & $   0.108$  & $   0.114$  & $   0.138$  & $   0.141$  & $   0.161$  & $   0.160$ \\
 $\Delta$C$_{\rm FIR, BI}$$^a$       & $\la$0.050 & $\la$0.050 & $\la$0.050 & $\la$0.050 & $\la$0.050 & $\la$0.050 & $\la$0.050 \\
\hline
\end{tabular}
\begin{flushleft}
$^a$ In fact, it also increases with metallicity.
\end{flushleft}
\label{Tab:dc-bi}
\end{table*}

\begin{table*}
\centering
\caption{Differences in the SFR conversion coefficients between at the lowest and highest metallicities $\Delta$C$_{\rm case, Z}$, $\Delta$C$_{\rm case,Z}$/$\Delta$[Fe/H] (the second column) and the variation rates of the SFR conversion coefficients with metallicity dC$_{\rm case, Z}$/d[Fe/H] (except C$_{\rm FIR}$, from the third to the last columns) for Models A, B, C-S55/Cha03, D-S55/K93', E-S55/K93' and F-S55'/K01 (only for the case of C$_{\rm H\alpha}$).
The top, second, third and bottom parts are for those in terms of $L_{\rm H\alpha}$, $L_{\rm 1500}$, $L_{\rm 2800}$ and $L_{\rm FIR}$, respectively. 
In each part (except the bottom one), two sub-parts are included.
The top and bottom sub-parts are for Models A-B and C-F/E, respectively, and the corresponding metallicity ranges are given in the first line (from the third to the last columns)}.
\begin{tabular}{ll llc ccr}

\hline
\hline
Model  &  $\Delta$C$_{\rm H\alpha, Z}$, {$\Delta \rm C_{\rm H\alpha, Z} \over \Delta[Fe/H]$} & \multicolumn{6}{c}{${\rm dC_{\rm H\alpha, Z} \over d[Fe/H]}$} \\
       &  (dex) & \multicolumn{6}{c}{} \\
\hline
  &  & [Fe/H]:$-$2.3$\sim$$-$1.8 & $-$1.8$\sim$$-$1.3 &$ -$1.3$\sim$$-$0.7 & $-$0.7$\sim$$-$0.3  &  $-$0.3$\sim$ 0.0  & 0.0$\sim$0.2 \\
   A &   0.4537, 0.181 &   0.0870 &   0.1158 &   0.1573 &   0.2577 &   0.3590 &   0.2355 \\
   B &   0.5171, 0.207 &   0.1106 &   0.1282 &   0.1963 &   0.2675 &   0.4253 &   0.2265 \\
  & \multicolumn{7}{l}{-----------------------------------------------------------------------------------------------------------------------------------------} \\
   &  & [Fe/H]:$-$2.3$\sim$$-$1.7 & $-$1.7$\sim$$-$0.7 & $-$0.7$\sim$$-$0.4 & $-$0.4$\sim$ 0.0  & 0.0$\sim$ 0.4  &   0.4$\sim$ 0.7 \\
   C-S55 &   0.4486, 0.166 &   0.1743 &   0.0646 &   0.1467 &   0.2378 &   0.3507 &   ...  \\
 C-Cha03 &   0.4488, 0.166 &   0.1742 &   0.0615 &   0.1440 &   0.2400 &   0.3590 &   ...  \\
   D-S55 &   0.5434, 0.259 &   ...    &   0.1019 &   0.2253 &   0.3600 &   0.5748 &   ...  \\
   D-K93'&   0.5426, 0.258 &   ...    &   0.0994 &   0.2240 &   0.3638 &   0.5762 &   ...  \\
   E-S55 &   0.4344, 0.145 &   0.1313 &   0.0811 &   0.2120 &   0.1845 &   0.3722 &  $-$0.0393 \\
   E-K93'&   0.4358, 0.145 &   0.1463 &   0.0976 &   0.2173 &   0.1900 &   0.3515 &  $-$0.1047 \\
   F-S55'&   0.4939, 0.183$^{\rm UN}$  &   0.0332$^{\rm UN}$ &   0.1974 &   0.2037 &   0.2513 &   0.2875&   ...  \\
   F-K01 &   0.6092, 0.226$^{\rm UN}$  &   0.0962$^{\rm UN}$ &   0.2060 &   0.3353 &   0.3470 &   0.2653 &   ...  \\

\hline
\hline
Model  &  $\Delta$C$_{\rm 1500, Z}$, {$\Delta \rm C_{\rm 1500, Z} \over \Delta[Fe/H]$} & \multicolumn{6}{c}{${\rm dC_{\rm 1500, Z} \over d[Fe/H]}$} \\
       &  (dex) & \multicolumn{6}{c}{} \\
\hline
  &  & [Fe/H]:$-$2.3$\sim$$-$1.8 & $-$1.8$\sim$$-$1.3 &$ -$1.3$\sim$$-$0.7 & $-$0.7$\sim$$-$0.3  &  $-$0.3$\sim$ 0.0  & 0.0$\sim$0.2 \\
        A&   0.2845,0.114 &   0.0600 &   0.0684 &   0.1213 &   0.1435 &   0.1960 &   0.1565\\
       B&   0.3065,0.123 &   0.0636 &   0.0728 &   0.1335 &   0.1535 &   0.2163 &   0.1595\\
 &  \multicolumn{7}{l}{-----------------------------------------------------------------------------------------------------------------------------------------} \\
   &  & [Fe/H]:$-$2.3$\sim$$-$1.7 & $-$1.7$\sim$$-$0.7 & $-$0.7$\sim$$-$0.4 & $-$0.4$\sim$ 0.0  & 0.0$\sim$ 0.4  &   0.4$\sim$ 0.7 \\
    C-S55  &   0.2363,0.088 &   0.0743 &   0.0639 &   0.0687 &   0.1428 &   0.1253 &  ...       \\
 C-Cha03&   0.2213,0.082 &   0.0683 &   0.0595 &   0.0590 &   0.1377 &   0.1200 &  ...       \\
   D-S55 &   0.2192,0.104 &  ...      &   0.0661 &   0.1240 &   0.1323 &   0.1575 &    ...     \\
   D-K93' &   0.2078,0.099 & ...      &   0.0609 &   0.1170 &   0.1235 &   0.1560 &   ...      \\
   E-S55 &   0.2490,0.083&   0.0398 &   0.0535 &   0.1000 &   0.1170 &   0.1765 &   0.0807\\
   E-K93'&   0.3322,0.111 &   0.0757 &   0.0885 &   0.1670 &   0.1740 &   0.2045 &  $-$0.0107\\

\hline
\hline
Model  &  $\Delta$C$_{\rm 2800, Z}$, {$\Delta \rm C_{\rm 2800, Z} \over \Delta[Fe/H]$} & \multicolumn{6}{c}{${\rm dC_{\rm 2800, Z} \over d[Fe/H]}$} \\
       &  (dex) & \multicolumn{6}{c}{} \\
\hline
  &  & [Fe/H]:$-$2.3$\sim$$-$1.8 & $-$1.8$\sim$$-$1.3 &$ -$1.3$\sim$$-$0.7 & $-$0.7$\sim$$-$0.3  &  $-$0.3$\sim$ 0.0  & 0.0$\sim$0.2 \\
       A&    0.3339,0.134$^{\rm UN}$&   0.0660$^{\rm UN}$ &   0.1240 &   0.1558 &   0.1695 &   0.1887 &   0.1050\\
       B&    0.3752,0.150$^{\rm UN}$&   0.0760$^{\rm UN}$ &   0.1406 &   0.1730 &   0.1867 &   0.2127 &   0.1230\\
 &  \multicolumn{7}{l}{-----------------------------------------------------------------------------------------------------------------------------------------} \\
   &  & [Fe/H]:$-$2.3$\sim$$-$1.7 & $-$1.7$\sim$$-$0.7 & $-$0.7$\sim$$-$0.4 & $-$0.4$\sim$ 0.0  & 0.0$\sim$ 0.4  &   0.4$\sim$ 0.7 \\
    C-S55&   0.2928,0.108$^{\rm UN}$ &   0.1050$^{\rm UN}$ &   0.1311 &   0.0850 &   0.1053 &   0.0777 &     ...    \\
 C-Cha03& 0.2648,0.098$^{\rm UN}$ &   0.0955$^{\rm UN}$ &   0.1202 &   0.0707 &   0.0955 &   0.0698 &   ...      \\
   D-S55&   0.2148,0.102$^{\rm UN}$ &    ...      &   0.0951$^{\rm UN}$ &   0.1443 &   0.1055 &   0.0855 &   ...      \\
   D-K93'&   0.1965,0.094$^{\rm UN}$ &    ...      &   0.0876$^{\rm UN}$ &   0.1333 &   0.0915 &   0.0808 &  ...       \\
   E-S55 &   0.3058,0.102$^{\rm UN}$ &   0.0872$^{\rm UN}$ &   0.1036 &   0.1257 &   0.1047 &   0.1235 &   0.0697\\
   E-K93'&   0.4521,0.151$^{\rm UN}$ &   0.1300$^{\rm UN}$ &   0.1653 &   0.2267 &   0.1897 &   0.1725 &  $-$0.0137\\

\hline
\hline
Model  &  $\Delta$C$_{\rm FIR, Z}$, {$\Delta \rm C_{\rm FIR, Z} \over \Delta[Fe/H]$} & \multicolumn{6}{c}{} \\
       &  (dex) & \multicolumn{6}{r}{} \\
       \hline
  A-B & $\sim$0.3, $\sim$0.1& \multicolumn{6}{r}{} \\
  C-F & $\sim$0.3-0.6, $\sim$0.1-0.2& \multicolumn{6}{r}{} \\

\hline
\end{tabular}
\label{Tab:dc-feh}
\end{table*}

\begin{table*}
\centering
\caption{Similar to Table~\ref{Tab:c-lha}, but for C$_{\rm 1500}$ and for Models A, B, C-S55/Cha03, D-S55/K93' and E-S55/K93'. The top and bottom parts are for Models A-B and C-E, respectively.}
\begin{tabular}{lcc ccc cc}

\hline
\hline
  Model & \multicolumn{7}{c}{C$_{\rm 1500}$,  $\sigma_{\rm 1500}$}\\
\hline
        & $Z=$0.0001 & $Z=$0.0003 & $Z=$0.001 & $Z=$0.004 & $Z=$0.01 & $Z=$0.02 &$Z=$0.03 \\
          A & $ -28.323,   0.043$  & $ -28.293,   0.035$  & $ -28.258,   0.030$  & $ -28.186,   0.016$  & $ -28.128,   0.012$  & $ -28.069,   0.009$  & $ -28.038,   0.008$ \\
         B & $ -28.278,   0.038$  & $ -28.246,   0.031$  & $ -28.210,   0.027$  & $ -28.130,   0.013$  & $ -28.068,   0.009$  & $ -28.003,   0.007$  & $ -27.972,   0.007$ \\
 
 & \multicolumn{7}{l}{------------------------------------------------------------------------------------------------------------------------------------------------------------------} \\
        & \ \ \ $Z=$0.0001 & $Z$=0.0004 & $Z$=0.004 & $Z$=0.008 & $Z$=0.02 & $Z$=0.05 & $Z$=0.1 \\
     C-S55 & $ -28.146,   0.044$  & $ -28.101,   0.038$  & $ -28.037,   0.028$  & $ -28.017,   0.026$  & $ -27.960,   0.025$  & $ -27.910,   0.024$  &                     \\
    C-Cha03 & $ -28.333,   0.043$  & $ -28.292,   0.037$  & $ -28.233,   0.028$  & $ -28.215,   0.026$  & $ -28.160,   0.025$  & $ -28.112,   0.023$  &                     \\
    D-S55 &              & $ -28.054,   0.021$  & $ -27.987,   0.009$  & $ -27.950,   0.007$  & $ -27.897,   0.005$  & $ -27.834,   0.004$  &                     \\
     D-K93' &              & $ -28.217,   0.019$  & $ -28.156,   0.008$  & $ -28.121,   0.006$  & $ -28.072,   0.005$  & $ -28.009,   0.004$  &                     \\
     E-S55 & $ -28.124,   0.026$  & $ -28.101,   0.018$  & $ -28.047,   0.008$  & $ -28.017,   0.006$  & $ -27.970,   0.004$  & $ -27.900,   0.003$  & $ -27.875,   0.031$ \\
     E-K93' & $ -28.083,   0.043$  & $ -28.038,   0.030$  & $ -27.949,   0.013$  & $ -27.899,   0.009$  & $ -27.830,   0.006$  & $ -27.748,   0.004$  & $ -27.751,   0.066$ \\

\hline
\end{tabular}
\label{Tab:c-luv1}
\end{table*}

\begin{table*}
\centering
\caption{Similar to Table~\ref{Tab:c-lha}, but for C$_{\rm 2800}$ and  for Models A, B, C-S55/Cha03, D-S55/K93' and E-S55/K93'. The top and bottom parts are for Models A-B and C-E, respectively.}
\begin{tabular}{lcl ccc cc}

\hline
\hline
  Model & \multicolumn{7}{c}{C$_{\rm 2800}$,  $\sigma_{\rm 2800}$}\\
\hline

        & $Z=$0.0001 & \ \ \ $Z=$0.0003 & $Z=$0.001 & $Z=$0.004 & $Z=$0.01 & $Z=$0.02 &$Z=$0.03 \\
         A & $ -28.333,   0.119^{\rm UN}$  & $ -28.300,   0.101$  & $ -28.238,   0.078$  & $ -28.145,   0.046$  & $ -28.077,   0.028$  & $ -28.020,   0.020$  & $ -27.999,   0.017$ \\
         B & $ -28.320,   0.124^{\rm UN}$  & $ -28.282,   0.102$  & $ -28.211,   0.078$  & $ -28.108,   0.046$  & $ -28.033,   0.027$  & $ -27.969,   0.020$  & $ -27.944,   0.016$ \\
  & \multicolumn{7}{l}{------------------------------------------------------------------------------------------------------------------------------------------------------------------} \\
        & $Z=$0.0001 & \ \ \ $Z$=0.0004 & $Z$=0.004 & $Z$=0.008 & $Z$=0.02 & $Z$=0.05 & $Z$=0.1 \\
     C-S55 & $ -28.138,   0.111^{\rm UN}$  & $ -28.075,   0.091$  & $ -27.944,   0.046$  & $ -27.918,   0.037$  & $ -27.876,   0.031$  & $ -27.845,   0.027$  &                     \\
C-Cha03 & $ -28.307,   0.105^{\rm UN}$  & $ -28.250,   0.086$  & $ -28.130,   0.044$  & $ -28.108,   0.036$   & $ -28.070,   0.030$  & $ -28.042,   0.026$  &                     \\
             D-S55 &              & $ -28.004,   0.081^{\rm UN}$  & $ -27.909,   0.037$  & $ -27.865,   0.023$  & $ -27.823,   0.012$  & $ -27.789,   0.009$  &                     \\
             D-K93' &              & $ -28.156,   0.076^{\rm UN}$  & $ -28.068,   0.034$  & $ -28.028,   0.021$  & $ -27.991,   0.012$  & $ -27.959,   0.008$  &                     \\
     E-S55 & $ -28.124,   0.106^{\rm UN}$  & $ -28.072,   0.081$  & $ -27.968,   0.032$  & $ -27.931,   0.019$  & $ -27.889,   0.011$  & $ -27.839,   0.006$  & $ -27.818,   0.034$ \\
     E-K93' & $ -28.177,   0.151^{\rm UN}$  & $ -28.099,   0.122$  & $ -27.934,   0.053$  & $ -27.865,   0.033$  & $ -27.790,   0.019$  & $ -27.721,   0.011$  & $ -27.725,   0.068$ \\

\hline
\end{tabular}
\label{Tab:c-luv2}
\end{table*}

In this section, we will use the Yunnan EPS models with and without binary interactions, various SFR forms and the algorithms described in Paper I, present the luminosity of H$\alpha$ recombination line $L_{\rm H\alpha}$, the luminosity of [OII]$\lambda$3727$\rm \AA$ forbidden line doublet $L_{\rm [OII]}$, the UV fluxes at 1500 and 2800\,$\rm \AA$, $L_{\rm i,UV}$, and FIR flux $L_{\rm FIR}$ of burst, E, S0, Sa-Sd and Irr galaxies, then present the SFR calibrations in terms of these diagnostics at $Z=0.0001, 0.0003, 0.001, 0.004, 0.01, 0.02$ and 0.03.
For the sake of clarity, we refer to those using the Yunnan models with and without binary interactions as Models A and B, respectively.

Using these two sets of models, we will discuss the effects of binary interactions and metallicity on the SFR calibrations in terms of $L_{\rm H\alpha}$, $L_{\rm [OII]}$, $L_{\rm i,UV}$ and $L_{\rm FIR}$.

\subsection{SFR versus $L_{\rm H\alpha}$, SFR versus $L_{\rm {[OII]}}$}
\label{Sect:sfr-Lha}
In Fig.~\ref{Fig:sfr-lha}, we give the relation between log(SFR) and log($L_{\rm H\alpha}$) (note the logarithmic scale) of E, S0, Sa-Sc and Sd galaxy types in the range from 0.1Myr to 15Gyr for Models A and B. For the sake of clarity, only the results at metallicity $Z$=0.0001, 0.001, 0.01 and 0.03 are presented. Also shown are the SFR($L_{\rm H\alpha}$) calibrations of K98 (${\rm SFR_{\rm H\alpha}}/{\rm M_\odot yr^{-1}}$=$7.9 \times 10^{-42} L_{\rm H\alpha}{\rm /erg\,s^{-1}}$) and \citet[][hereafter B04, ${\rm SFR_{\rm H\alpha}}/{\rm M_\odot yr^{-1}}$=$5.25 \times 10^{-42} L_{\rm H\alpha}{\rm /erg\,s^{-1}}$]{bri04}. Both of calibration relations are linear and obtained at solar metallicity.

From Fig.~\ref{Fig:sfr-lha}, we see that
the SFR($\rm H\alpha$) calibration curves of all galaxy types for Models A and B at different metallicities are parallel to those of K98 and B04. That is to say, SFR varies linearly with $L_{\rm H\alpha}$ for all galaxy types and at all $Z$. 
For each set of models, the SFR($\rm H\alpha$) calibration lines of all galaxy types overlap at a given $Z$, and the SFR($\rm H\alpha$) calibration line moves upwards with increasing metallicity for a given galaxy type.
At $Z=0.03$, both sets of SFR($\rm H\alpha$) calibration lines for Models A and B locate above that of K98, which is above that of B04 by $\sim$\,0.2\,dex. Why the calibration lines move upwards with increasing $Z$? When increasing $Z$, the temperature of stars would decrease, hence the UV flux, the number of ionizing photons $Q$(H) and $L_{\rm H\alpha}$ (see equation \ref{Eq:lha}) would decrease.
Comparing the results between Models A and B at a given metallicity, we can obtain the effect of binary interactions on the SFR($L_{\rm H\alpha}$) calibration.
From Fig.~\ref{Fig:sfr-lha}, we see that the set of SFR($\rm H\alpha$) calibration curves of Model A locates below that of Model B at all $Z$, and the distance from the set of SFR($\rm H\alpha$) calibration curves of Model A to Model B increases with $Z$.
The reason is that binary interactions can produce more hotter stars at all $Z$, these hotter stars have relatively significant contribution (although smaller $f_{\rm BS}$ and $f_{\rm HeMS}$, see Table~\ref{Tab:rst-bi}) to the UV flux and $L_{\rm H\alpha}$ at high $Z$ because that the temperature of stars in SPs would decrease with increasing $Z$.

To quantitatively analyze the effects of binary interactions and metallicity on the conversion coefficient between SFR and $L_{\rm H\alpha}$, we give a fitting relation between log(SFR) and log($L_{\rm H\alpha}$) at a given $Z$ when log(SFR)$\ge -11$ and $|$log(SFR)$-$log(SFR)$_{0}|$$>$0.05 [log(SFR)$_{0}$ is the value of the corresponding galaxy type at $t$=0.1\,Myr, approximately equal to the log(SFR) value of horizontal line in Fig.~\ref{Fig:sfr-lha}] for Models A and B by the following form:
\begin{equation}
{\rm log } {\rm SFR_{\rm H\alpha} \over (\rm M_\odot \ {\rm yr^{-1}})} =
{\rm log} {L_{\rm H\alpha} \over (\rm erg \ s^{-1})} + {\rm
C_{H\alpha}},
\label{Eq:ha-c1}
\end{equation}
where SFR$_{\rm H\alpha}$ means that it is calculated from $L_{\rm H\alpha}$. This form of fitting (equation~\ref{Eq:ha-c1}) is the same as that of K98 and MPD98. In Table~\ref{Tab:c-lha}, we present the fitting coefficient (${\rm C_{H\alpha}}$) and the rms ($\sigma_{\rm H{\alpha}}$) for all models at all metallicities, the results of Models A and B are presented in the second and the third lines.

From Table~\ref{Tab:c-lha}, we see that
${\rm C_{H\alpha}}$ increases and $\sigma_{\rm H{\alpha}}$ shows little change when increasing $Z$ for both Models A and B.
The inclusion of binary interactions makes C$_{\rm H\alpha}$ smaller at all metallicities and the effect of binary interactions on the C$_{\rm H\alpha}$ increases with $Z$, this also can be seen from the first line of Table~\ref{Tab:dc-bi}, in which we give the differences in the SFR conversion coefficients between Models A and B, $\Delta$C$_{\rm case, BI}$ (=$|{\rm C_{case,A}}$$-$${\rm C_{case,B}}|$), at different metallicities. The inclusion of binary interactions makes C$_{\rm H{\alpha}}$ smaller by $\sim$0.1\,dex at $Z$=10$^{-3}$ and $\sim$0.2\,dex at $Z$=0.03.

Comparing the conversion coefficient at different metallicities for a given set of models, we can obtain the effect of metallicity on the SFR calibration.
In the second column of Table~\ref{Tab:dc-feh}, we give the differences in the SFR conversion coefficients $\Delta$C$_{\rm case,Z}$ (=$ |{\rm C_{case, zmax}} - {\rm C_{case, zmin}}|$) between at the highest and lowest metallicities (zmax, zmin) and $\Delta$C$_{\rm case,Z}$/$\Delta$[Fe/H] [[Fe/H] means metallicity expressed by the iron abundance relative to the Sun, =log($Z/Z_\odot$)] for all models. 
From the second and the third lines of the top part, we can that the values of Model A are less than the corresponding ones of Model B, i.e. the inclusion of binary interactions lowers the sensitivity of C$_{\rm H\alpha}$ to metallicity (caused by the fact that $\Delta$C$_{\rm H\alpha, BI}$ increases with $Z$).
However, the variation rate of SFR conversion coefficient with metallicity (dC$_{\rm case, Z}$/d[Fe/H]) is different within different metallicity ranges, thus we give them in the 3rd-8th columns of Table~\ref{Tab:dc-feh}    for all models. From the second and the third lines of the top part, we see that d${\rm C_{H\alpha,Z}}$/d[Fe/H] reaches the maximal value near solar metallicity ($\sim$0.36 and $\sim$0.43) for Models A and B.

In this work, the luminosity of [OII]$\lambda$3727$\rm \AA$ forbidden line doublet, $L_{\rm [OII]}$, is obtained by using the empirical ratio $L_{\rm {[OII]}}/L_{\rm H\alpha}$=0.23, which is used in the work of \citet{hop03}.
Because the fixed $L_{\rm {[OII]}}/L_{\rm H\alpha}$ ratio is used, the SFR($L_{\rm [OII]}$) calibration curve moves upwards by an amount of lg(1/0.23) in comparison with that of SFR($L_{\rm H\alpha}$) in Fig.~\ref{Fig:sfr-lha}.
The effects of binary interactions and metallicity on the SFR($L_{\rm [OII]}$) calibration are the same as those on the SFR($L_{\rm H\alpha}$) calibration. In the fourth line of Table~\ref{Tab:dc-bi}, we give the difference in  the SFR($L_{\rm [OII]}$) calibration factor, C${_{\rm [OII]}}$, between Models A and B at different metallicities.

\subsection{SFR versus $L_{\rm 1500}$ and SFR versus $L_{\rm 2800}$}
\label{Sect:sfr-Luv}
In Fig.~\ref{Fig:sfr-luv}, we give the relations between log(SFR) and the logarithmic UV luminosities at 1500 and 2800\,$\rm \AA$ of E, S0-Sd types of galaxies for Models A and B. For the sake of clarity, only the results at $Z$=0.0001 and 0.03 are presented in Fig.~\ref{Fig:sfr-luv}. 
The relations between log(SFR) and log($L_{i, {\rm UV}}$) at other metallicities lie between those at $Z$=0.0001 and 0.03.
Also shown are the results of K98, \citet[][hereafter MPD98]{mad98} and \citet[][hereafter G10]{gil10}. K98 gives ${\rm SFR_{\rm UV}}/{\rm M_\odot yr^{-1}}$=1.4$\times 10^{-28} L_{i, {\rm UV}} / {\rm erg\,s^{-1}\,Hz^{-1}}$, MPD98 gives ${\rm SFR_{\rm UV}}/{\rm M_\odot yr^{-1}}$=C$\times 10^{-28} L_{i,\rm UV} / {\rm erg\,s^{-1}\,Hz^{-1}}$, where C=(1.25, 2.86) for $L_{\rm 1500}$ and C=(1.26, 1.96) for $L_{\rm 2800}$ when using the IMFs of S55 and \citet{sca86}, and G10 gives ${\rm SFR_{\rm UV}}/{\rm M_\odot yr^{-1}}$ = 0.71$\times 10^{-28} L_{i, {\rm UV}} / {\rm erg\,s^{-1}\,Hz^{-1}}$.

First, from the left-hand panel of Fig.~\ref{Fig:sfr-luv}, we see that
at $Z=0.03$, $L_{\rm 1500}$ varies linearly with SFR for all galaxy types (in comparison with the lines of K98, MPG98 and G10) and the SFR($L_{\rm 1500}$) calibration curves of all galaxy types (from E to Sd) overlap (coincide)  for both Models A and B (i.e. a good SFR indicator at high $Z$). Two sets of SFR($L_{\rm 1500}$) calibration lines locate below those of K98 and MPD98 with the S55 IMF ($\sim$0.1\,dex) and above that of G10 ($\sim$0.1\,dex). 
At $Z$=0.0001, the SFR($L_{\rm 1500}$) calibration curves of all galaxy types do not display the same calibration relation (out of alignment/not in a line, the difference of $\sim$0.4\,dex), only those of late types display the linear relation for Models A and B.
The earlier (i.e. $\tau$ decreases) is the galaxy type, the more is the deviation from the linear SFR-$L_{1500}$ relation (the larger is the slope of calibration curve) and the lower is the location of the SFR($L_{1500}$) calibration curve (not a good SFR indicator at low $Z$).
At both metallicities, the inclusion of binary interactions can lower the SFR($L_{1500}$) conversion coefficient.
The set of SFR($L_{1500}$) calibration lines moves upwards when $Z$ is from 0.0001 to 0.03 for Models A and B.

From the right-hand panel of Fig.~\ref{Fig:sfr-luv}, we see that the SFR($L_{\rm 2800}$) calibration is similar to that of SFR($L_{\rm 1500}$) for Models A and B, but there are two exceptions. One is at $Z$=0.03, the SFR($L_{\rm 2800}$) calibration is not unique for all galaxy types (the difference of less than 0.05\,dex), while is unique for the SFR($L_{\rm 1500}$) calibration.
The second is at $Z$=0.0001, the calibration curves do not display the linear calibration relation for all galaxy types, while only for early types in the case of $L_{\rm 1500}$.
Moreover, the deviation from the linear SFR-$L_{\rm 2800}$ relation (slope $>$ 1) and the deviation from the unique SFR-$L_{\rm 2800}$ relation among all galaxy types are far larger than those in the case of $L_{\rm 1500}$. This phenomenon means that $L_{2800}$ can not be used in the linear calibration of SFR at low-metallicity end.

Also, for the purpose of quantitative analyse, in Tables~\ref{Tab:c-luv1} and ~\ref{Tab:c-luv2}, we give the linear fitting coefficients (similar to equation 8, C$_{1500}$ and C$_{2800}$) and the rms ($\sigma_{1500}$ and $\sigma_{2800}$) between log(SFR) and log($L_{1500}$) and between log(SFR) and log($L_{2800}$) for all models when log(SFR)$>$$-11.0$ and $|$log(SFR)$-$log(SFR)$_{0}|$$>$0.05 at a given metallicity. 
Because the relation between $L_{2800}$ and SFR at low $Z$ is not unique and not linear for all galaxy types, the coefficients related to $L_{2800}$ at $Z=10^{-3}$ ($\Delta$C$_{\rm 2800, BI}$ at $Z=10^{-3}$,  $\Delta$C$_{\rm 2800, Z}$, $\Delta {\rm C_{2800, Z}} \over \Delta {\rm [Fe/H]}$,  ${\rm dC_{2800, Z}} \over {\rm d[Fe/H]}$ within the [Fe/H] range from $-2.3$ to $-1.8$, C$_{\rm 2800}$ at $Z=10^{-3}$) are labelled by a superscript of 'UN' in Tables~\ref{Tab:dc-bi}, \ref{Tab:dc-feh} and \ref{Tab:c-luv2} for Models A and B.
From the second and the third lines of Tables~\ref{Tab:c-luv1} and ~\ref{Tab:c-luv2}, we see that C$_{\rm 1500}$ and C$_{\rm 2800}$ increase and $\sigma_{\rm 1500}$ and $\sigma_{\rm 2800}$ decrease when increasing $Z$ for both Models A and B. The relative large rms at low metallicities is caused by the non-unique relation among all galaxy types and the non-linear relation between log(SFR) and log($L_{i,\rm UV}$) for both Models A and B.
The inclusion of binary interactions lowers C$_{i, {\rm UV}}$ at all metallicities.

From the differences in the C$_{\rm 1500}$ and C$_{\rm 2800}$ between Models A and B ($\Delta$C$_{\rm 1500, BI}$ and $\Delta$C$_{\rm 2800, BI}$) at different metallicities, which are presented in the second and the third lines of Table~\ref{Tab:dc-bi}, we see that the effect of binary interactions on the C$_{\rm 1500}$ and C$_{\rm 2800}$ increases with $Z$ (0.045 $\rightarrow$ 0.066\,dex for C$_{\rm 1500}$ and 0.013 $\rightarrow$ 0.055\,dex for C$_{\rm 2800}$). This phenomenon also can be seen from the comparison in the distance from the set of SFR($L_{i,{\rm UV}}$) lines of Models A to B between at $Z$=0.0001 and 0.03 in Fig.~\ref{Fig:sfr-luv}, but this conclusion is prone to be affected by the larger rms at low metallicities.

At last, from the second and the third lines in the second and third parts of Table~\ref{Tab:dc-feh}, we see again that the values of Model A are less than the corresponding ones of Model B. This also means that the inclusion of binary interactions lowers the sensitivity of the SFR($L_{\rm 1500}$) and SFR($L_{\rm 2800}$) calibrations to metallicity.
The dC$_{\rm 1500, Z}$/d[Fe/H] and dC$_{\rm 2800, Z}$/d[Fe/H] are different within different metallicity ranges and also reach the maximal value near solar metallicity ($\sim$0.20) for Models A and B.

\subsection{SFR versus $L_{\rm FIR}$}
For the SFR calibrations in terms of $L_{\rm FIR}$, it is from the models with constant SFR under the assumption of the bolometric luminosity $L_{\rm BOL}$=$L_{\rm FIR}$.

In Fig.~\ref{Fig:sfr-lfir}, we give the $L_{\rm FIR}$ evolution of Irr galaxies (i.e. models with constant SFR) for Models A and B. For the sake of clarity, only the results at $Z$=0.0001, 0.001, 0.01 and 0.03 are presented. Also shown are the result of K98 (${\rm SFR_{\rm FIR}}/{\rm M_\odot yr^{-1}} = 4.5 \times 10^{-44} L_{\rm FIR}/{\rm erg\,s^{-1}}$).
From it, we see that SFR does not vary linearly with $L_{\rm FIR}$ for Models A and B at all metallicities. The lower is the metallicity, the larger is the deviation from the linear SFR-$L_{\rm FIR}$ relation.

Moreover, from Fig.~\ref{Fig:sfr-lfir}, we see that the inclusion of binary interactions raises the $L_{\rm FIR}$, thus lowers the SFR($L_{\rm FIR}$) conversion factor C$_{\rm FIR}$ at all metallicities (less than 0.05\,dex, see the fifth line of Table~\ref{Tab:dc-bi}). From the distance from the SFR($L_{\rm FIR}$) calibration line of Models A to B in Fig.~\ref{Fig:sfr-lfir}, we can see that it increases with metallicity, i.e. the effect of binary interactions on the SFR($L_{\rm FIR}$) calibration increases with metallicity.
Furthermore, for the above reason, $\Delta$C$_{\rm FIR}$/$\Delta$[Fe/H] of Model A is less than that of Model B, i.e. the inclusion of binary interactions decreases the sensitivity of C$_{\rm FIR}$ to metallicity.

At last, the SFR($L_{\rm FIR}$) calibration coefficient increases with metallicity. The effect of metallicity on the SFR($L_{\rm FIR}$) calibration factors, $\Delta$C$_{\rm FIR}$, and ${{\Delta {\rm C_{FIR}}}\over {\Delta {\rm [Fe/H]}}}$ reaches $\sim$0.3\,dex and 0.1 (see the first line of the bottom part of Table 3).
Because it is difficult to get the exact linear calibration coefficient C$_{\rm FIR}$, in Table 3 we do not give the variation rate of C$_{\rm FIR}$ with metallicity, dC$_{\rm FIR}$/d[Fe/H], within different metallicity ranges.

\subsection{Comments}
Moreover, from Table~\ref{Tab:dc-feh}, we see that $\Delta$C$_{\rm H\alpha}$/$\Delta$[Fe/H], $\Delta$C$_{\rm 2800}$/$\Delta$[Fe/H] and $\Delta$C$_{\rm 1500}$/$\Delta$[Fe/H] decrease in succession for Models A and B. This means that SFR($L_{\rm H\alpha}$) calibration is the most sensitive to metallicity than those of SFR($L_{\rm 2800}$) and SFR($L_{\rm 1500}$) (in turn).

\section{SFR calibrations by using the other EPS models}
\begin{figure}
\centering
\includegraphics[height= 8.0cm,width=6.5cm,angle=270]{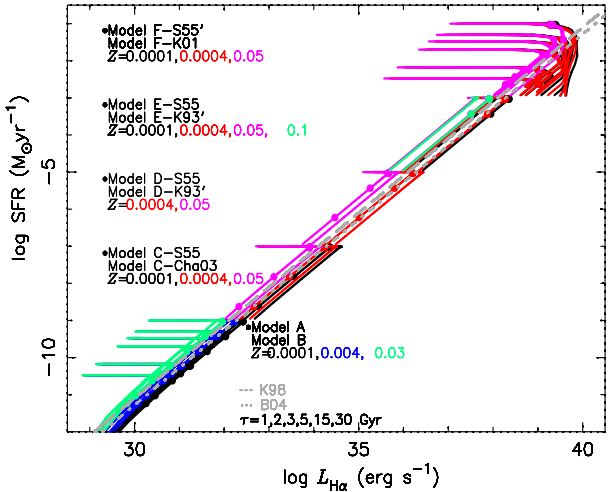}
\caption{Relation between SFR and $L_{\rm H\alpha}$ for Models A (solid circles), B, C-S55 (circles), C-Cha03, D-S55 (circles), D-K93', E-S55 (circles), E-K93', F-S55' (circles) and F-K01 at different metallicities. For Models C-F, the results are moved upwards along the diagonal line, respectively.
For Models A/B and F-S55'/K01, E, S0 and Sa-Sd types (from top to bottom) are included, while for Models C-S55/Cha03, D-S55/K93' and E-S55/K93', only E type is included.
For Models A/B, $Z$=0.0001 (black), 0.004 (blue) and 0.03 (green, from right to left) are included,
for Models C-S55/Cha03 and F-S55'/K01, $Z$=0.0001 (black), 0.0004 (red) and 0.05 (magenta) are included,
for Models D-S55/K93', $Z$=0.0004 (red) and 0.05 (magenta) are included, and
for Models E-S55/K93', $Z$=0.0001 (black), 0.0004 (red), 0.05 (magenta) and 0.1 (green) are included.
Note the color is the same at a given $Z$.
At last, also shown are the results of K98 (grey dashed line) and B04 (grey dotted line).}
\label{Fig:sfr-lha-com}
\end{figure}

\begin{figure*}
\includegraphics[angle=270,scale=.410]{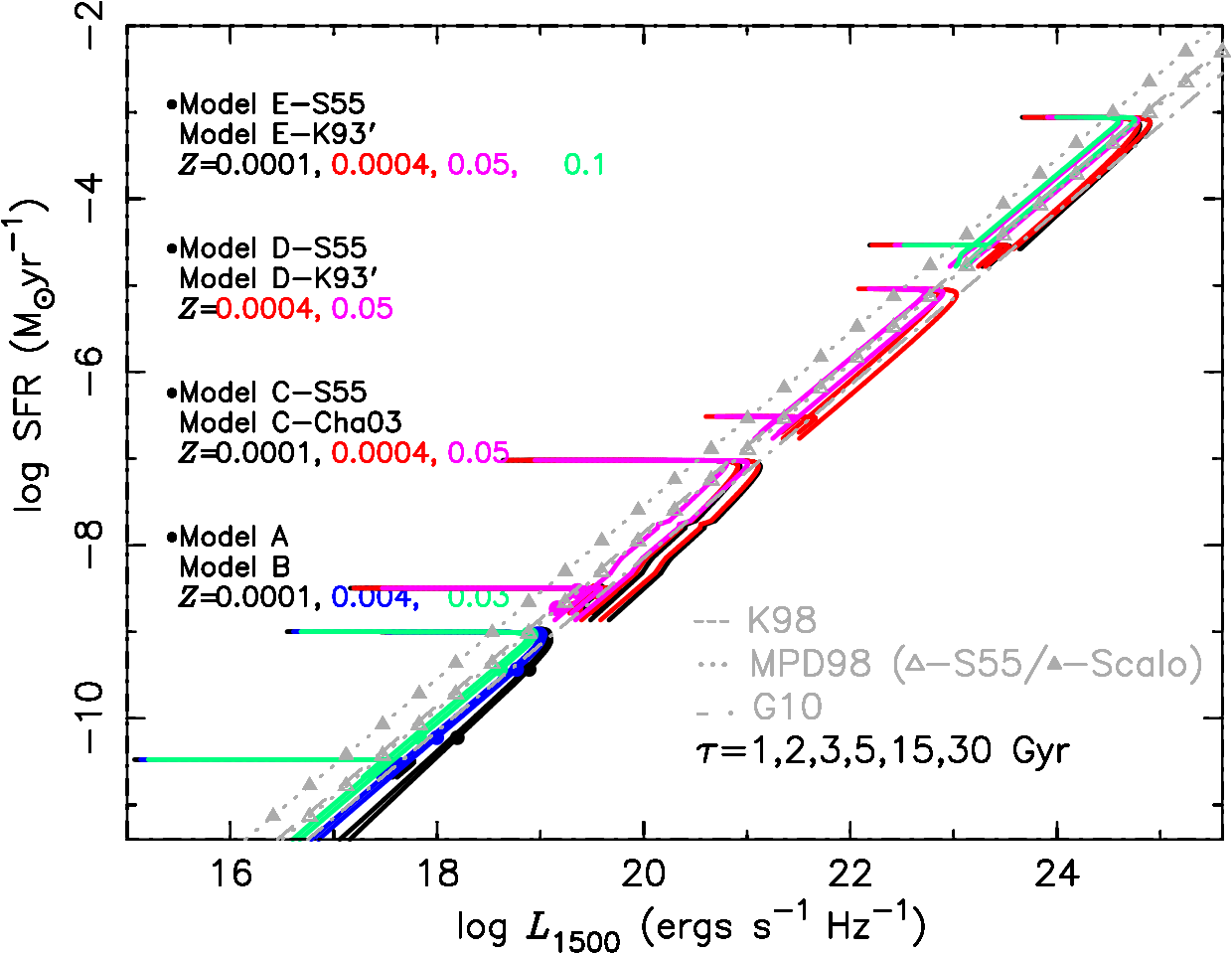}
\includegraphics[angle=270,scale=.410]{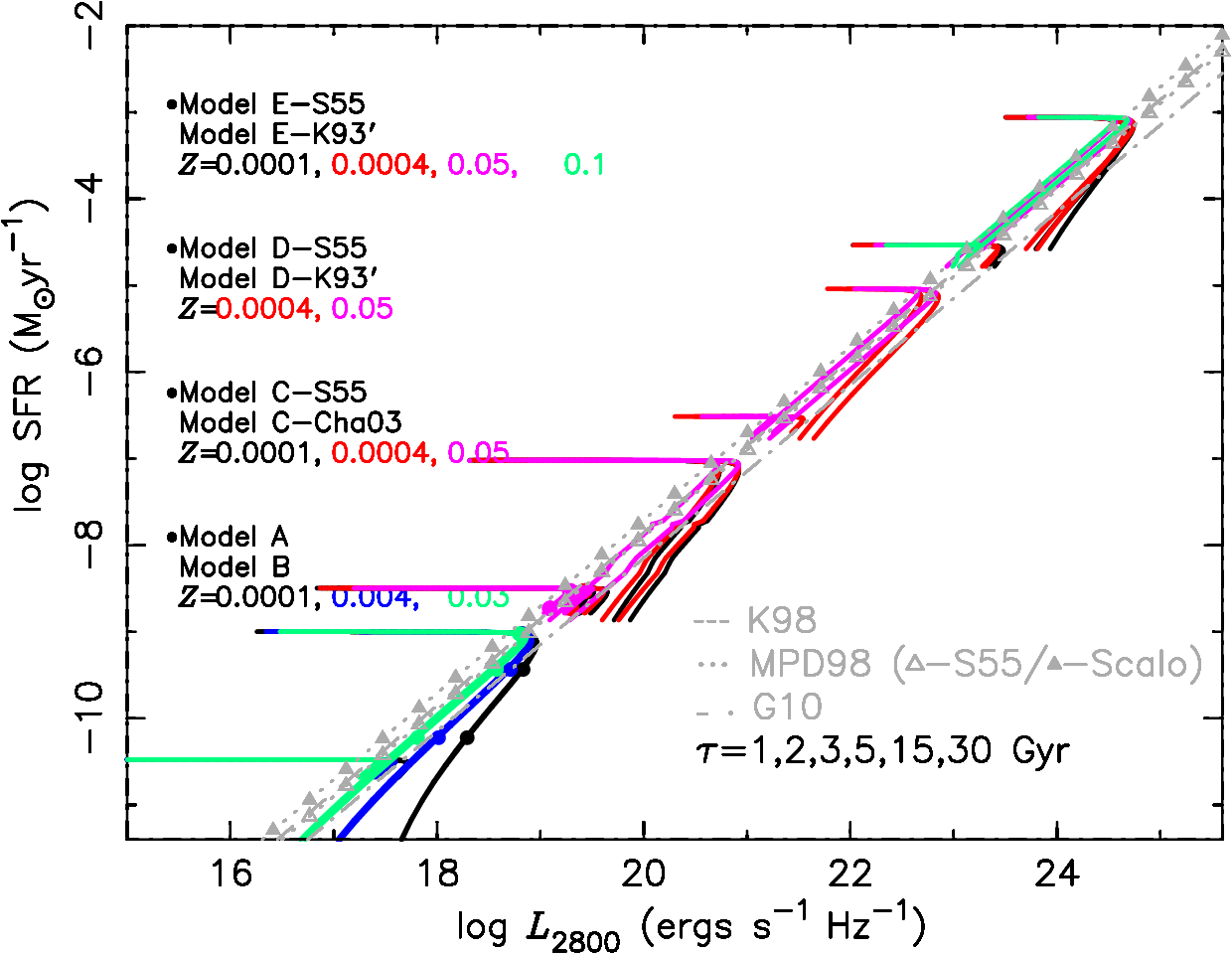}
\caption{Relations between SFR and $L_{i, \rm UV}$ for Models A/B, C-S55/C-Cha03, D-S55/D-K93' and E-S55/E-K93' at different metallicities. Left-hand panel is for $L_{1500}$ and right-hand panel is for $L_{2800}$.
For each set of models, the selected metallicities are the same as those in Fig.~\ref{Fig:sfr-lha-com}, and only E and Irr galaxy types are presented. The calibration line colour (representing metallicity) and symbol (IMF) have the same meanings as in Fig.~\ref{Fig:sfr-lha-com}. For Models C-E, the results are moved upwards along the diagonal line, respectively.
Also shown are the results of K98 (grey dashed line), MPD98 (grey dotted line, open and solid triangles are for using the S55 and Scalo IMFs, respectively) and G10 (grey dot-dashed line).}
\label{Fig:sfr-luv-com}
\end{figure*}

\begin{figure}
\centering
\includegraphics[height= 8.5cm,width= 6.5cm,angle=270]{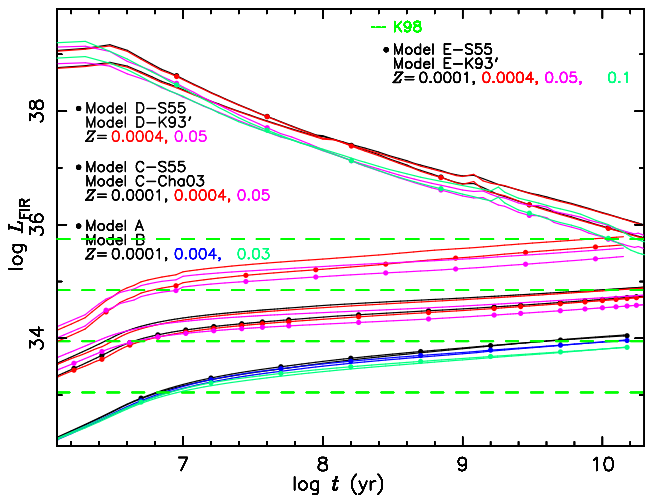}
\caption{The $L_{\rm FIR}$ evolution of Irr galaxies (i.e. models with constant star formation, SFR=1\,M$_{\rm \odot}$) for Models A/B, C-S55/C-Cha03, D-S55/D-K93' and E-S55/E-K93'.
The line colour and symbol have the same meanings as in Fig.~\ref{Fig:sfr-lha-com}. Also shown is the result of K98 (green dashed line).
The results of Models C-E and K98 are moved upwards, respectively.}
\label{Fig:fir-t-com}
\end{figure}

\begin{table*}
\centering
\caption{Differences in the SFR conversion coefficients between obtained by using the two IMFs, $\Delta$C$_{\rm case,IMF}$, for Models C, D, E and F (only for $\Delta$C$_{\rm H\alpha}$) at metallicity $Z$=0.0001, 0.0004, 0.004, 0.008, 0.02, 0.05 and 0.1. The top, second, third and bottom parts are for the cases of $\Delta$C$_{\rm H\alpha}$, $\Delta$C$_{\rm1500}$, $\Delta$C$_{\rm 2800}$ and $\Delta$C$_{\rm FIR}$, respectively.}
\begin{tabular}{cll rrr rr}
\hline
Model & $Z$=0.0001  & 0.0004 & 0.004 & 0.008 & 0.02 & 0.05 & 0.1 \\
\hline
\multicolumn{8}{c}{$\Delta$C$_{\rm H\alpha, IMF}$ (dex)}\\
C &  $-$0.221&  $-$0.221&  $-$0.224&  $-$0.225&  $-$0.224&  $-$0.221& \\
D &        &  $-$0.198&  $-$0.200&  $-$0.201&  $-$0.199&  $-$0.199& \\
E & \ \ 0.299&   \ \ 0.308&   0.325&   0.326&   0.329&   0.320&   0.301\\
F &  $-$0.181$^{\rm UN}$&  $-$0.143&  $-$0.135&  $-$0.095&  $-$0.057&  $-$0.066& \\

\hline
\multicolumn{8}{c}{$\Delta$C$_{\rm 1500, IMF}$ (dex)}\\
C &  $-$0.187&  $-$0.191&  $-$0.196&  $-$0.198&  $-$0.200&  $-$0.202& \\
D &             &  $-$0.163&  $-$0.169&  $-$0.171&  $-$0.175&  $-$0.175&  \\
E &\ \ 0.041&  \  \ 0.063&   0.098&   0.118&   0.140&   0.152&   0.124\\

\hline
\multicolumn{8}{c}{$\Delta$C$_{\rm 2800, IMF}$ (dex)}\\
C &      $-$0.169$^{\rm UN}$ &  $-$0.175&  $-$0.186&  $-$0.190&  $-$0.194&  $-$0.197& \\
D & &  $-$0.152$^{\rm UN}$ &  $-$0.159&  $-$0.163&  $-$0.168&  $-$0.170&  \\
E &      $-$0.053$^{\rm UN}$ &  $-$0.027&   0.034&   0.066&   0.099&   0.118&   0.993\\

\hline
\multicolumn{8}{c}{$\Delta$C$_{\rm FIR, IMF}$ (dex)}\\
C-E$^{a}$ & $\le$0.3 & $\le$0.3 &$\le$0.3 &$\le$0.3 &$\le$0.3 &$\le$0.3 &$\le$0.3 \\

\hline
\end{tabular}
\begin{flushleft}
*$^a$ In fact, $\Delta$C$_{\rm case,IMF}$ is different for different models, and has no significant correlation with metallicity.
\end{flushleft}
\label{Tab:dc-imf}
\end{table*}

%
To discuss the effects of metallicity, EPS models and IMF on these SFR calibrations when using the other EPS models, compare the conclusions (about the effect of metallicity on these SFR calibrations) with those from the Yunnan models, in this section we will present the SFR($L_{\rm H\alpha}$), SFR($L_{\rm [OII]}$), SFR($L_{i, \rm UV}$) and SFR($L_{\rm FIR}$) calibrations by using the BC03, {\hf SB99}, {\hf P\'EGASE} and POPSTAR EPS models. These EPS models (including the used parameters, physics, IMFs, $M_{\rm l}$, $M_{\rm u}$ and metallicities) have been described in Section 2.
These four sets of results are referred to as Models C, D, E and F, respectively.
For each set of results, two subsets are considered, depending on the IMF. To distinguish them, the name of used IMF is the supplement to the model name (see the first column of Table~\ref{Tab:mod-des}).

Using the above EPS models at their own metallicities (see the fifth column of Table~\ref{Tab:mod-des}), in this section, we first obtain the $L_{\rm H\alpha}$, $L_{\rm [OII]}$, $L_{i, \rm UV}$ and $L_{\rm FIR}$ of burst, E, S0, Sa-Sd and Irr types of galaxies, then give the linear fitting coefficients between log(SFR) and log($L_{\rm H\alpha}$), between log(SFR) and log($L_{\rm 1500}$) and between log(SFR) and log($L_{\rm 2800}$) in Tables~\ref{Tab:c-lha}, ~\ref{Tab:c-luv1} and~\ref{Tab:c-luv2}, and their variation rates with metallicity in the upper, second and third parts of Table~\ref{Tab:dc-feh}.
Because the calibration between SFR and $L_{\rm FIR}$ is not linear, we only give the $\Delta$C$_{\rm FIR}$ and $\Delta$C$_{\rm FIR}$/$\Delta$[Fe/H] in the bottom part of Table~\ref{Tab:dc-feh}.

\subsection{SFR versus $L_{\rm H\alpha}$}
\label{Sect:sfr-Lha-com}
First, we study the relation between SFR and $L_{\rm H\alpha}$ for Models C-S55/Cha03, D-S55/K93', E-S55/K93' and F-S55'/K01, and find that SFR varies linearly with $L_{\rm H\alpha}$ [the slope d(logSFR)/d(log$L_{\rm H\alpha}$)$\sim$1] for all galaxy types (E-Sd), metallicities and models except the F-S55'/K01 models (SFR varies linearly with $L_{\rm H\alpha}$ only for all galaxy types at high metallicities, while not true for early galaxy types at low metallicities). The SFR($L_{\rm H\alpha}$) calibration lines of all galaxy types overlape for a given set of models with $Z$.
Therefore, for the sake of clarity, in Fig.~\ref{Fig:sfr-lha-com} we only give the relation between log(SFR) and log($L_{\rm H\alpha}$) of E, S0, Sa, Sb, Sc and Sd types for Models A/B at $Z$=0.0001, 0.004 and 0.03 and for Models F-S55'/K01 at $Z$=0.0001, 0.0004 and 0.05, that of only E type for Models C-S55/Cha03 at $Z$=0.0001, 0.0004 and 0.05, for Models D-S55/K93' at $Z$=0.0004 and 0.05 and for Models E-S55/K93' at $Z$=0.0001, 0.0004, 0.05 and 0.1.
The reason we choose these metallicities (highlighted in red in Table~\ref{Tab:mod-des}) for a given set of models is that the metallicity value is either the upper/lower limit or the common one.
Moreover, for the sake of clarity, the log(SFR) versus log($L_{\rm H\alpha}$) calibration curves of  Models C-F in Fig.~\ref{Fig:sfr-lha-com} are moved upwards along the diagonal line [i.e. log(SFR) and log($L_{\rm H\alpha}$) are multiplied by the same factor for a given set of models].
At last, we also give the results of K98 and B04 in Fig.~\ref{Fig:sfr-lha-com}.

From Fig.~\ref{Fig:sfr-lha-com}, we see exactly that
the slope dlog(SFR)/dlog($L_{\rm H\alpha}$) is similar to those of K98 and B04 for all galaxy types, all models and all metallicities (except for early types of Models F-S55'/K01 at low metallicities), i.e. SFR varies linearly with $L_{\rm H\alpha}$. 
For Models F-S55'/K01 at low metallicities, the slope of SFR($L_{\rm H\alpha}$) calibration curves of all galaxy types deviates from those of K98 and B04 (mainly for early types) and the calibration lines of all galaxy types do not overlape. 
Therefore, the values related to $L_{\rm H\alpha}$ at $Z$=$10^{-3}$ in Tables~\ref{Tab:c-lha},  \ref{Tab:dc-feh} and \ref{Tab:dc-imf} (C$_{\rm H\alpha}$ at $Z$=$10^{-3}$, $\Delta$C$_{\rm H\alpha, Z}$, $\Delta{\rm C_{H\alpha, Z}} \over \Delta{\rm[Fe/H]}$, ${\rm dC_{H\alpha, Z}} \over {\rm d[Fe/H]}$ in the [Fe/H] range from $-$2.3 to $-$1.7 and $\Delta$C$_{\rm H\alpha, IMF}$ at $Z$=$10^{-3}$) for Models F-S55'/K01 are labelled by a superscript of 'UN'. For a given set of models, the set of SFR($L_{\rm H\alpha}$)
calibration curves moves upwards with increasing metallicity, i.e. the conversion factor increases (this also can be seen from the values in Table~\ref{Tab:c-lha}), but there are an exception: Models E-S55/K93' in the [Fe/H] range from 0.4 to 0.7. The value of dC$_{\rm H\alpha,Z}$/d[Fe/H] is negative within this range for Models E-S55/K93' (see the top panel of Table~\ref{Tab:dc-feh}). The reason that C$_{\rm H\alpha}$ increases with $Z$ is that the temperature of stars decreases with increasing $Z$.

{\bf [Fe/H]:} From the top part of Table~\ref{Tab:dc-feh}, we see that 
dC$_{\rm H\alpha, Z}$/d[Fe/H] is different within different [Fe/H] ranges and reaches the maximum value near  solar metallicity for all models.
Comparing the values of all models, we find that $\Delta$C$_{\rm H\alpha, Z}$/$\Delta$[Fe/H] and the maximum value of dC$_{\rm H\alpha, Z}$/d[Fe/H] are the largest (dC$_{\rm H\alpha, Z}$/d[Fe/H]$\sim$0.57 in the [Fe/H] range from 0.0 to 0.4) for Models D-S55/K93', i.e. Models D-S55/K93' are the most sensitive to metallicity for the SFR($L_{\rm H\alpha, Z}$) calibration.

{\bf IMF:} In Table~\ref{Tab:dc-imf}, we give the differences in the SFR conversion coefficients between obtained by using the two IMFs, $\Delta$C$_{\rm case,IMF}$ (=C$_{\rm case,Cha03/K93'/K01} -$ C$_{\rm case,S55/S55'}$), for Models C-F/E at different metallicities. 
From the top part, we see that $\Delta$C$_{\rm H\alpha, IMF}$ is independent of $Z$ for Models C, D and E ($\sim$$-$0.2, $\sim$$-$0.2 and $\sim$0.3\,dex, respectively), the absolute value of $\Delta$C$_{\rm H\alpha, IMF}$ for Model F decreases with increasing $Z$ ($\sim$0.18$\rightarrow$0.07\,dex from $Z$=0.0001 to 0.05). 
The larger $|$$\Delta$C$_{\rm H\alpha, IMF}$$|$ at low metallicities for Model F partly is caused by the deviation from the linear SFR-$L_{\rm H\alpha}$ relation.
At last, it can be seen that $|\Delta$C$_{\rm H\alpha, IMF}|$ of Model E is greater than those of 
Models C, D and F, i.e. Model E is the most sensitive to IMF for the SFR($L_{\rm H\alpha, Z}$) calibration.

{\bf EPS and other:} From Table~\ref{Tab:c-lha}, we see that the difference in the C$_{\rm H\alpha}$ reaches $\sim$0.55\,dex at $Z$=0.02 among Models A-F (the seventh and the sixth columns for Models A-B and C-F) and $\sim$0.52\,dex at $Z$=0.0001 among Models A-C and E (the second column, excluding Models F-S55'/K01 because of the non-linear SFR-$L_{\rm H\alpha}$ relation). This kind of difference is comparable to that caused by metallicity, and mainly is caused by the differences in the adoption of EPS models, the algorithm of obtaining $L_{\rm H\alpha}$ and IMF.

Excluding the results from the models considering binary interactions or those using the K93'/K01/Cha03 IMFs, from Table~\ref{Tab:c-lha}, we see that the difference in the C$_{\rm H\alpha}$ is $\sim$0.20\,dex among Models B, C-S55, D-S55, E-S55 and F-S55' at $Z$=0.02 and $\sim$0.02\,dex among Models B, C-S55 and E-S55 at $Z$=0.0001. 
These differences (less than 0.2\,dex) are mainly caused by the adoptions of different EPS models and the algorithm of obtaining $L_{\rm H\alpha}$.

For Models B-C and D, the algorithm of obtaining $L_{\rm H\alpha}$ and the companied coefficients (see equation \ref{Eq:lha}) are the same, so the difference in the C$_{\rm H\alpha}$ among these models ($\sim$0.184\,dex at $Z$=0.02; $\sim$0.02\,dex at $Z$=$10^{-3}$) is caused by EPS models.
For Models E-S55/K93', the algorithm of obtaining $L_{\rm H\alpha}$ is different from that of Models A-D. For Models F-S55'/K01, the algorithm is the same but the coefficients depend on $Z$ and electronic temperature.

\subsection{SFR versus $L_{1500}$ and SFR versus $L_{2800}$}
\label{Sect:sfr-Luv-com}
Here, we also check the relations between log(SFR) and log($L_{i, \rm UV}$) for Models A, B, C-S55/Cha03, D-S55/K93' and E-S55/K93' in advance, and find that SFR does not linearly vary with $L_{i, \rm UV}$ in some cases (for example, SFR versus $L_{\rm 2800}$ for early types of Models A-E at low metallicities).
For a set of models with a given $Z$, the SFR($L_{i, \rm UV}$) calibration curves of all galaxy types are not unique in some cases  [such as SFR($L_{\rm 2800}$) for Models A-E at low metallicities].
Therefore, for the sake of clarity, in Fig.~\ref{Fig:sfr-luv-com} we give the relations between log(SFR) and log($L_{i, \rm UV}$) of only E and Irr galaxy types for Models A, B, C-S55/Cha03, D-S55/K93' and E-S55/K93'. For a given set of models, the selected metallicities are the same as those in Fig.~\ref{Fig:sfr-lha-com}. Similarly, the results of Models C, D and E are moved upwards along the diagonal line. In Fig.~\ref{Fig:sfr-luv-com}, we also give the results of K98, MPD98 and G10.

From the left- and right-hand panels of Fig.~\ref{Fig:sfr-luv-com}, we see that the calibration curves of SFR versus $L_{1500}$ and SFR versus $L_{2800}$ for Models C, D and E are similar to the corresponding ones for Models A and B.
SFR varies linearly with $L_{1500}$ and $L_{2800}$ [the slope log(SFR)/log($L_{i, \rm UV}$) $\sim$ 1, in comparison with those of K98, MPD98 and G10] for all models, all galaxy types and all metallicities except for early types at low metallicities (more significant for $L_{2800}$).
And at low metallicities ($Z$=0.0001 or 0.0004), the SFR($L_{i, \rm UV}$) calibration is not unique for all galaxy types, for example,  that of SFR($L_{\rm 1500}$) at low metallicities for Models A-C and that of SFR($L_{\rm 2800}$) at low metallicities for all models (more significant). 
Due to the above facts, the values ($\Delta \rm C_{\rm 2800,Z}$, $\Delta{\rm C_{2800,Z}} \over \Delta{\rm [Fe/H]}$, ${\rm dC_{2800,Z}} \over {\rm d[Fe/H]}$, C$_{\rm 2800}$ and $\Delta$C$_{\rm 2800, IMF}$) 
related to  $L_{\rm 2800}$ at low metallicities ($Z$=$10^{-3}$ for Models C and E, $Z$=4$\times10^{-3}$ for Model D) in Tables~\ref{Tab:dc-feh}, \ref{Tab:c-luv2} and \ref{Tab:dc-imf} are labelled by a superscript of 'UN' for all models.
The conversion coefficients C$_{i, {\rm UV}}$ increase with increasing $Z$ for all models  (except Model E-K93' in the [Fe/H] range of 0.4-0.7). The linear fitting coefficients, between SFR and $L_{1500}$ and between SFR and $L_{2800}$, are presented in Tables~\ref{Tab:c-luv1} and \ref{Tab:c-luv2}, respectively.

{\bf [Fe/H]:} 
From the second and third parts of Table~\ref{Tab:dc-feh}, we see that 
dC$_{\rm 1500, Z}$/d[Fe/H] and dC$_{\rm 2800, Z}$/d[Fe/H] are different within different [Fe/H] ranges and reach the maximum value near solar metallicity (similar to that of dC$_{\rm H\alpha}$/d[Fe/H]) and at low metallicities for all models, respectively.
Among all models, $\Delta$C$_{\rm 1500/2800, Z}$/$\Delta$[Fe/H] and the maximal values of dC$_{\rm 1500/2800, Z}$/d[Fe/H] are the largest for Model E-K93', i.e. Model E-K93' is the most sensitive to metallicity for the SFR($L_{\rm 1500/2800, Z}$) calibrations.

{\bf IMF:} From the second and third parts of Table~\ref{Tab:dc-imf}, we see that $\Delta$C$_{\rm 1500,IMF}$ and $\Delta$C$_{\rm 2800,IMF}$ are independent of $Z$ for Models C and D ($\sim$\,$-$0.2\,dex and $\sim$$-$0.17\,dex, respectively). The absolute $\Delta$C$_{\rm 1500,IMF}$ and $\Delta$C$_{\rm 2800,IMF}$ increase with increasing $Z$ for Model E (difference of $\sim$0.1\,dex), this is different from the case of $\Delta$C$_{\rm H\alpha,IMF}$ for Model E (independent of $Z$).
Moreover, among all models, the difference in the C$_{i, {\rm UV}}$ caused by IMF is the largest for Model C at all metallicities, i.e. Model C is the most sensitive to IMF for the SFR($L_{i, {\rm UV}}$) calibrations.

{\bf EPS:} From the SFR($L_{1500}$) and SFR($L_{2800}$) calibration curves in Fig.~\ref{Fig:sfr-luv-com} and the conversion coefficients in Tables~\ref{Tab:c-luv1} and \ref{Tab:c-luv2} for all models, we see that the differences in the C$_{1500}$ and C$_{2800}$ are $\sim$0.36 and 0.28\,dex at $Z$=0.02 for Models A-E (the seventh and sixth columns for Models A-B and C-E) and  $\sim$0.25 and 0.2\,dex at $Z$=0.0001 among Models A-C and E (the second column, comparable to those caused by metallicity). The above differences are mainly caused by the differences in the IMF and the adoption of EPS models.
From Tables~\ref{Tab:c-luv1} and ~\ref{Tab:c-luv2}, we see that the differences in the C$_{1500}$ and C$_{2800}$ are $\sim$0.11 and 0.15\,dex among Models B, C-S55 D-S55 and E-S55 at $Z$=0.02 and $\sim$0.13 and 0.14\,dex at $Z$=0.0001 among Models B, C-S55 and E-S55, so the adoption of EPS models causes to the difference of $\sim$0.2\,dex in the C$_{1500}$ and C$_{2800}$.

\subsection{SFR versus $L_{\rm FIR}$}
\label{Sect:sfr-Lfir-com}
In Fig.~\ref{Fig:fir-t-com}, we give the evolution of bolometric magnitude of Irr galaxies for Models A, B, C, D and E. For a given set of models, the selected metallicities are the same as those in Fig.~\ref{Fig:sfr-lha-com}. Also shown are the result of K98. Similarly, the results of Models C, D and E and the result of K98 are moved upwards.

From Fig.~\ref{Fig:fir-t-com}, we see that the effect of metallicity on the SFR versus $L_{\rm FIR}$ calibration for Models C-E is similar to that for Models A-B: C$_{\rm FIR}$ increases with increasing $Z$, the difference in the SFR($L_{\rm FIR}$) conversion factor caused by metallicity and $\Delta$C$_{\rm FIR}$/$\Delta$[Fe/H] reach $\sim$0.3-0.6\,dex and 0.1-0.2 (see the last line of Table 3).
The difference in the C$_{\rm FIR}$ caused by EPS models reaches $\sim$1.2\,dex (excluding Model E), which is two/more times larger than that caused by metallicity.
The difference in the C$_{\rm FIR}$ caused by IMF is insignificant (less than 0.3\,dex, see the last line of Table 8) in comparison with that caused by EPS models. The difference caused by IMF seems to be independent of metallicity.

\subsection{Comments}
In total, for a given set of models, the value of $\Delta$C$_{\rm H\alpha}$/$\Delta$[Fe/H] is lager than those of $\Delta$C$_{\rm 2800,Z}$/$\Delta$[Fe/H] and $\Delta$C$_{\rm 1500,Z}$/$\Delta$[Fe/H] in turn.
i.e. C$_{\rm H\alpha}$ is the most sensitive to metallicity than those of C$_{2800}$ and C$_{1500}$ (the lines are densely concentrated)
However, there are some exceptions. For Model D, $\Delta$C$_{\rm 1500,Z}$/$\Delta$[Fe/H] $>$$\Delta$C$_{\rm 2800,Z}$/$\Delta$[Fe/H]. For Model E-K93', $\Delta$C$_{\rm 2800,Z}$/$\Delta$[Fe/H] $>$ $\Delta$C$_{\rm H\alpha}$/$\Delta$[Fe/H].

\section{Comparison among various factors and the implications}
\begin{table}
\centering
\caption{Differences in the SFR conversion coefficients, $\Delta$C$_{\rm case, factor}$, caused by the adoption of different EPS models, metallicity and IMF.}
\begin{tabular}{lrc r} 
\hline
Case & EPS  & $Z$ & IMF \\
\hline
$\Delta$C$_{\rm H\alpha, factor}$ (dex) & $\sim$0.2 & 0.43-0.61   & 0.06-0.33  \\
$\Delta$C$_{\rm 1500, factor}$ (dex)      & $\sim$0.2 & 0.20-0.33   & 0.04-0.20 \\
$\Delta$C$_{\rm 2800, factor}$ (dex)      & $\sim$0.2 & 0.20-0.45   & 0.03-0.20 \\
$\Delta$C$_{\rm FIR, factor}$ (dex)        & $\sim$1.2 & $\sim$0.3-0.6   & $\le$0.3  \\
\hline
\end{tabular}
\label{Tab:dc-fact}
\end{table}

\subsection{Effects of various factors on these SFR calibrations}
In this part, we will summary and compare the effects of binary interactions, metallicity, EPS models and IMF on the SFR conversion coefficients in terms of $L_{\rm H\alpha}$, $L_{\rm [OII]}$, $L_{i, {\rm UV}}$ and $L_{\rm FIR}$.

From Section 4, we see that the inclusion of binary interactions lowers all SFR conversion factors (C$_{\rm H\alpha}$, C$_{\rm [OII]}$, C$_{i, {\rm UV}}$ and C$_{\rm FIR}$) considered in this paper, the differences in these conversion factors caused by the inclusion of binary interactions increase with $Z$ (less than 0.2\,dex, see Table 4). 
Moreover, the inclusion of binary interactions lowers the sensitivity of these SFR calibrations to metallicity.

From Sections 4 and 5, we see that the SFR calibration coefficients increase with metallicity in general, the maximal value of dC$_{\rm case,Z}$/d[Fe/H] is near solar metallicity or at low metallicity ranges.
Among all models, Models D and E-K93' are the most sensitive to metallicity for the C$_{\rm H\alpha}$ and C$_{i, {\rm UV}}$, respectively.

From Section 5, we see that the differences in the C$_{\rm H\alpha}$, C$_{\rm 1500}$, C$_{\rm 2800}$ and C$_{\rm FIR}$ among Models A-E/F reach $\sim$0.5, $\sim$0.3, $\sim$0.3 and $\sim$1.5\,dex at a given metallicity. However, after excluding the effects of IMF and algorithm, the differences in the C$_{\rm H\alpha}$, C$_{\rm 1500}$, C$_{\rm 2800}$ and C$_{\rm FIR}$, which are solely caused by EPS models, are $\sim$0.2, $\sim$0.2, $\sim$0.2 and $\sim$1.2\,dex.

The differences in these conversion coefficients caused by the adoption of IMF (see Table 8) are independent of metallicity for Models C-F except for $\Delta$C$_{\rm H\alpha, IMF}$ of Model F (the absolute value decreases with $Z$) and $\Delta$C$_{i, {\rm UV, IMF}}$ of Model E (the absolute value increases with $Z$). 
Among all models, the effect of IMF on the C$_{\rm H\alpha}$ and C$_{i, {\rm UV}}$ is the largest (the most sensitive to IMF) for Models E and C, respectively.

At last, from Sections 4 and 5, we see that C$_{\rm H\alpha}$, C$_{2800}$, C$_{1500}$ are less sensitive to metallicity in turn for a given set of models.

In Table~\ref{Tab:dc-fact}, we summary the differences in these conversion coefficients caused by the adoption of different EPS models, metallicity and IMF. The values in the third and fourth columns of Table~\ref{Tab:dc-fact} are from Tables~\ref{Tab:dc-feh} and \ref{Tab:dc-imf}.
From it, we see that the difference in the C$_{\rm H\alpha}$ caused by metallicity is two times larger than that caused by the adoption of EPS models and IMF.
The difference in the SFR($L_{\rm FIR}$) calibration factors caused by the adoption of EPS models is two/more times larger than that caused by metallicity/IMF.
As for the SFR($L_{i, {\rm UV}}$) calibration, the effects of EPS models, IMF and metallicity are comparable.

K98 has even summarized that the effects of metallicity and IMF on the C$_{\rm UV}$ and C$_{\rm H\alpha}$ reach to $\sim$0.3 and $\sim$0.1\, dex. From Table~\ref{Tab:dc-fact}, we see that the maximal differences in the C$_{i, {\rm UV}}$ and C$_{\rm H\alpha}$ caused by metallicity and IMF are several times larger than those presented by K98.

\subsection{Effects of binary interactions and metallicity on the discrepancy in the SFR}
In Fig. 1 of \citet{hop04}, we see that the SFR derived from $L_{\rm H\alpha}$ (SFR$_{\rm H\alpha}$) is larger than that from the UV luminosity (SFR$_{\rm UV}$) for star-forming galaxies in the redshift range of $z \la 6$. In his work, all SFRs are obtained by using the linear calibrations of K98 at solar metallicity and have been corrected by IMF (i.e. using the S55 IMF).

Using the definition of the linear SFR calibration factor (see equation 8), the following formula can be derived,
\begin{equation}
 {\rm SFR_{A, 2} \over SFR_{B, 2}} =  {\rm SFR_{A, 1} \over SFR_{B, 1}} {\rm 10^{(C_{A,2}-C_{A,1}) - (C_{B,2}-C_{B,1})}},
\end{equation}
where A and B mean that SFR is derived from $L_{\rm H\alpha}$ and $L_{\rm i,UV}$, and '1' and '2' correspond to the first (or standard) and second cases.

{\bf (i)} If '1' and '2' correspond to the cases of solar and low metallicities, from the values in Tables 3, 6 and 7, we can get ${\rm (C_{A,2}-C_{A,1}) > (C_{B,2}-C_{B,1})}$,  i.e. ${\rm SFR_{A, 2} \over SFR_{B, 2}} >  {\rm SFR_{A, 1} \over SFR_{B, 1}}$.
This implies that the difference between SFR$_{\rm H\alpha}$ and SFR$_{\rm UV}$, displayed in Fig. 1 of \citet{hop04}, will be enhanced if using the calibration coefficients at low metallicities.

{\bf (ii)} If '1' and '2' are in the cases of neglecting and including binary interactions at solar metallicity, also from the values in Tables 3, 6 and 7, we can get ${\rm (C_{A,2}-C_{A,1}) > (C_{B,2}-C_{B,1})}$. This also implies that the difference between SFR$_{\rm H\alpha}$ and SFR$_{\rm UV}$ will be enhanced if using the calibration coefficients when considering binary interactions.

\section{Summary and conclusions}
Using the Yunnan EPS models with and without binary interactions, we present the $L_{\rm H\alpha}$, $L_{\rm [OII]}$, $L_{i, {\rm UV}}$ and $L_{\rm FIR}$ for burst, E, S0, Sa-Sd and Irr galaxies, the conversion coefficients between SFR and these diagnostics at $Z=$0.0001, 0.0003, 0.001, 0.004, 0.01, 0.02 and 0.03, and discuss the effects of binary interactions and metallicity (see the next paragraph) on these calibrations of SFR.
The inclusion of binary interactions lowers the SFR versus $L_{\rm H\alpha}$ and SFR versus $L_{\rm [OII]}$ conversion factors by $\sim$0.1-0.2\,dex, the SFR versus $L_{\rm 1500}$ by $\sim$0.055\,dex, the SFR versus $L_{\rm 2800}$ by $\sim$0.035\,dex and the SFR versus $L_{\rm FIR}$ by $\sim$0.05\,dex.
The differences in these conversion coefficients caused by the inclusion of binary interactions are dependent of metallicity.
The higher is the metallicity, the larger are the differences in these conversion factors.
The inclusion of binary interactions lowers the sensitivity of these SFR calibrations to metallicity.

We also obtain the $L_{\rm H\alpha}$, $L_{\rm [OII]}$, $L_{i, {\rm UV}}$ and $L_{\rm FIR}$ for burst, E, S0, Sa-Sd and Irr galaxies by using the BC03 (0.0001$\le Z \le$0.05), {\hf SB99} (0.0004 $\le Z \le$0.05), {\hf P\'EGASE} (0.0001$\le Z \le$0.1) and POPSTAR (0.0001$\le Z \le$0.05) models, and present the conversion coefficients between SFR and these diagnostics.
For these models, we discuss the effects of IMF, EPS models (see the next paragraph) and metallicity on these SFR calibrations, and compare the conclusions about the effect of metallicity on these SFR calibrations with those from the Yunnan EPS models.
By comparisons, we find that the conclusions are similar. 
For each set of models,  {\bf (i)} the relations between SFR and these diagnostics are linear for all galaxy types at all metallicities (except for $L_{\rm FIR}$, $L_{\rm H\alpha}$ when using the POPSTAR models and $L_{i, {\rm UV}}$ for early types at low metallicities when using any set of models);
{\bf (ii)} the $L_{i, {\rm UV}}$ (especially for $L_{\rm 2800}$) is not suitable to the linear calibration of SFR at low metallicities;
{\bf (iii)} the conversion coefficients between SFR and these tracers increase with $Z$ except C$_{\rm H\alpha}$ and C$_{i, {\rm UV}}$ when using the {\hf P\'EGASE} models within 0.4$\le$[Fe/H]$\le$0.7 (only Model E-K93' for C$_{i, {\rm UV}}$);
{\bf (iv)} the dC$_{\rm H\alpha,Z}$/d[Fe/H], dC$_{\rm 1500,Z}$/d[Fe/H] and dC$_{\rm 2800,Z}$/d[Fe/H] reach the maximum value near solar metallicity or at low metallicities, respectively;
{\bf (v)} the values of $\Delta{\rm C_{\rm H\alpha,Z}} \over \Delta{\rm [Fe/H]}$, $\Delta{\rm C_{\rm 2800,Z}} \over \Delta{\rm [Fe/H]}$ and $\Delta{\rm C_{\rm 1500,Z}} \over \Delta{\rm [Fe/H]}$ decrease in turn, it means that C$_{\rm H\alpha}$ is the most sensitive to metallicity (in general).
Among all models, Models D and E-K93' are the most sensitive to metallicity for the SFR($L_{\rm H\alpha}$)  and SFR($L_{i, {\rm UV}}$) calibrations.

The uncertainties in these SFR calibrations caused by EPS models and IMF are as follows. {\bf (i)} The differences in the SFR($L_{\rm H\alpha}$), SFR($L_{i,{\rm UV}}$) and SFR($L_{\rm FIR}$) calibration factors caused by the adoption of EPS models reach $\sim$0.2, 0.2 and 1.2\,dex.
{\bf (ii)} The differences in the SFR($L_{\rm H\alpha}$) and SFR($L_{i,{\rm UV}}$) conversion coefficients caused by IMF are in the range of 0.03-0.33\,dex (see Tables 8 and~\ref{Tab:dc-fact}) for all models, and these differences are independent of $Z$ for a given set of models (except $\Delta$C$_{\rm H\alpha, IMF}$ when using the POPSTAR models and $\Delta$C$_{i, {\rm UV, IMF}}$ when using the {\hf P\'EGASE} models).
Among all models, Models E and C are the most sensitive to IMF for the SFR($L_{\rm H\alpha}$) and SFR($L_{i, {\rm UV}}$) calibrations.

Comparing the discrepancies in these SFR calibrations caused by IMF, EPS models and metallicity, we find that the differences in the SFR($L_{\rm H\alpha}$) and SFR($L_{\rm FIR}$) calibrations are mainly caused by metallicity/IMF and EPS models, respectively. The effects of EPS models, metallicity (relatively large) and IMF on the C$_{i, {\rm UV}}$ are comparable.
Moreover, our derived differences in the SFR conversion factors caused by IMF and metallicity are larger than those mentioned by K98.
The difference between SFR$_{\rm H\alpha}$ and SFR$_{\rm UV}$ will be enlarged when using the calibration factors obtained at low metallicities or those obtained when considering binary interactions.

\section*{acknowledgements}
This work was funded by the Chinese Natural Science Foundation (Grant Nos 11273053, 11073049, 11033008, 10821026 \& 2007CB15406), by Yunnan Foundation (Grant No 2011CI053) and by the Chinese Academy of Sciences (KJCX2-YW-T24).
We are also grateful to the referee for suggestions that have improved the quality of this paper.

\bibliography{zfh-mn}

\begin{thebibliography}{50}
\expandafter\ifx\csname natexlab\endcsname\relax\def\natexlab#1{#1}\fi

\bibitem[{{Aarseth}(1999)}]{ara99}
{Aarseth} S.~J., 1999, \pasp, 111, 1333

\bibitem[{{Anders} {et~al}\mbox{.}(2012){Anders}, {Baumgardt}, {Gaburov}, \&
  {Portegies Zwart}}]{and12}
{Anders} P., {Baumgardt} H., {Gaburov} E., {Portegies Zwart} S., 2012, \mnras,
  421, 3557

\bibitem[{{Brinchmann} {et~al}\mbox{.}(2004){Brinchmann}, {Charlot}, {White},
  {Tremonti}, {Kauffmann}, {Heckman}, \& {Brinkmann}}]{bri04}
{Brinchmann} J., {Charlot} S., {White} S.~D.~M., {Tremonti} C., {Kauffmann} G.,
  {Heckman} T., {Brinkmann} J., 2004, \mnras, 351, 1151

\bibitem[{{Bruzual} \& {Charlot}(2003)}]{bru03}
{Bruzual} G., {Charlot} S., 2003, \mnras, 344, 1000

\bibitem[{{Chabrier}(2003)}]{cha03}
{Chabrier} G., 2003, \pasp, 115, 763

\bibitem[{{Crowther}(2012)}]{cro12}
{Crowther} P., 2012, Astronomy and Geophysics, 53, 040000

\bibitem[{{de Grijs} {et~al}\mbox{.}(2008){de Grijs}, {Goodwin}, {Kouwenhoven},
  \& {Kroupa}}]{deg08}
{de Grijs} R., {Goodwin} S.~P., {Kouwenhoven} M.~B.~N., {Kroupa} P., 2008,
  \aap, 492, 685

\bibitem[{{Eldridge}(2012)}]{eld12}
{Eldridge} J.~J., 2012, \mnras, 422, 794

\bibitem[{{Ferland}(1980)}]{fer80}
{Ferland} G.~J., 1980, \pasp, 92, 596

\bibitem[{{Fioc} \& {Rocca-Volmerange}(1997)}]{fio97}
{Fioc} M., {Rocca-Volmerange} B., 1997, \aap, 326, 950

\bibitem[{{Fioc} \& {Rocca-Volmerange}(1999)}]{fio99}
{Fioc} M., {Rocca-Volmerange} B., 1999, \aap, 344, 393

\bibitem[{{Gao} \& {Solomon}(2004)}]{gao04}
{Gao} Y., {Solomon} P.~M., 2004, \apjs, 152, 63

\bibitem[{{Gilbank} {et~al}\mbox{.}(2010){Gilbank}, {Baldry}, {Balogh},
  {Glazebrook}, \& {Bower}}]{gil10}
{Gilbank} D.~G., {Baldry} I.~K., {Balogh} M.~L., {Glazebrook} K., {Bower}
  R.~G., 2010, \mnras, 405, 2594

\bibitem[{{Hamann} \& {Koesterke}(1998)}]{ham98}
{Hamann} W.-R., {Koesterke} L., 1998, \aap, 335, 1003

\bibitem[{{Hern{\'a}ndez} \& {Bruzual}(2011)}]{her11}
{Hern{\'a}ndez} F.~C., {Bruzual} G., 2011, in Revista Mexicana de Astronomia y
  Astrofisica Conference Series, Vol.~40, Revista Mexicana de Astronomia y
  Astrofisica Conference Series, pp. 277--277

\bibitem[{{Hopkins}(2004)}]{hop04}
{Hopkins} A.~M., 2004, \apj, 615, 209

\bibitem[{{Hopkins} {et~al}\mbox{.}(2003){Hopkins}, {Miller}, {Nichol},
  {Connolly}, {Bernardi}, {G{\'o}mez}, {Goto}, {Tremonti}, {Brinkmann},
  {Ivezi{\'c}}, \& {Lamb}}]{hop03}
{Hopkins} A.~M. {et~al.}, 2003, \apj, 599, 971

\bibitem[{{Horiuchi} {et~al}\mbox{.}(2013){Horiuchi}, {Beacom}, {Bothwell}, \&
  {Thompson}}]{hor13}
{Horiuchi} S., {Beacom} J.~F., {Bothwell} M.~S., {Thompson} T.~A., 2013, ArXiv
  e-prints

\bibitem[{{Humphreys} \& {Davidson}(1994)}]{hum94}
{Humphreys} R.~M., {Davidson} K., 1994, \pasp, 106, 1025

\bibitem[{{Hurley} {et~al}\mbox{.}(2005){Hurley}, {Pols}, {Aarseth}, \&
  {Tout}}]{hur05}
{Hurley} J.~R., {Pols} O.~R., {Aarseth} S.~J., {Tout} C.~A., 2005, \mnras, 363,
  293

\bibitem[{{Hurley} {et~al}\mbox{.}(2002){Hurley}, {Tout}, \& {Pols}}]{hur02}
{Hurley} J.~R., {Tout} C.~A., {Pols} O.~R., 2002, \mnras, 329, 897

\bibitem[{{Kang} {et~al}\mbox{.}(2012){Kang}, {Zhang}, \& {Zhang}}]{kan12}
{Kang} X., {Zhang} F., {Zhang} Y., 2012, Science in China G: Physics and
  Astronomy, 55, 1505

\bibitem[{{Kennicutt}(1998)}]{ken98}
{Kennicutt}, Jr. R.~C., 1998, \araa, 36, 189

\bibitem[{{Kewley} {et~al}\mbox{.}(2003){Kewley}, {Geller}, \&
  {Jansen}}]{kew03}
{Kewley} L.~J., {Geller} M.~J., {Jansen} R.~A., 2003, in Bulletin of the
  American Astronomical Society, Vol.~35, American Astronomical Society Meeting
  Abstracts, p. 119.01

\bibitem[{{Kobulnicky} \& {Fryer}(2007)}]{kob07}
{Kobulnicky} H.~A., {Fryer} C.~L., 2007, \apj, 670, 747

\bibitem[{{Kouwenhoven} {et~al}\mbox{.}(2007){Kouwenhoven}, {Brown}, {Portegies
  Zwart}, \& {Kaper}}]{kou07}
{Kouwenhoven} M.~B.~N., {Brown} A.~G.~A., {Portegies Zwart} S.~F., {Kaper} L.,
  2007, \aap, 474, 77

\bibitem[{{Kroupa} {et~al}\mbox{.}(2001){Kroupa}, {Aarseth}, \&
  {Hurley}}]{kro01}
{Kroupa} P., {Aarseth} S., {Hurley} J., 2001, \mnras, 321, 699

\bibitem[{{Kroupa} {et~al}\mbox{.}(1993){Kroupa}, {Tout}, \& {Gilmore}}]{kro93}
{Kroupa} P., {Tout} C.~A., {Gilmore} G., 1993, \mnras, 262, 545

\bibitem[{{Leitherer}(2008)}]{lei08}
{Leitherer} C., 2008, in IAU Symposium, Vol. 255, IAU Symposium, {Hunt} L.~K.,
  {Madden} S.~C., {Schneider} R., eds., pp. 305--309

\bibitem[{{Leitherer} {et~al}\mbox{.}(2010){Leitherer}, {Ortiz Ot{\'a}lvaro},
  {Bresolin}, {Kudritzki}, {Lo Faro}, {Pauldrach}, {Pettini}, \& {Rix}}]{lei10}
{Leitherer} C., {Ortiz Ot{\'a}lvaro} P.~A., {Bresolin} F., {Kudritzki} R.-P.,
  {Lo Faro} B., {Pauldrach} A.~W.~A., {Pettini} M., {Rix} S.~A., 2010, \apjs,
  189, 309

\bibitem[{{Leitherer} {et~al}\mbox{.}(1999){Leitherer}, {Schaerer}, {Goldader},
  {Gonz{\'a}lez Delgado}, {Robert}, {Kune}, {de Mello}, {Devost}, \&
  {Heckman}}]{lei99}
{Leitherer} C. {et~al.}, 1999, \apjs, 123, 3

\bibitem[{{Madau} {et~al}\mbox{.}(1998){Madau}, {Pozzetti}, \&
  {Dickinson}}]{mad98}
{Madau} P., {Pozzetti} L., {Dickinson} M., 1998, \apj, 498, 106

\bibitem[{{Meynet} \& {Maeder}(2000)}]{mey00}
{Meynet} G., {Maeder} A., 2000, \aap, 361, 101

\bibitem[{{Miller} \& {Scalo}(1979)}]{mil79}
{Miller} G.~E., {Scalo} J.~M., 1979, \apjs, 41, 513

\bibitem[{{Moll{\'a}} {et~al}\mbox{.}(2009){Moll{\'a}}, {Garc{\'{\i}}a-Vargas},
  \& {Bressan}}]{mol09}
{Moll{\'a}} M., {Garc{\'{\i}}a-Vargas} M.~L., {Bressan} A., 2009, \mnras, 398,
  451

\bibitem[{{Nieuwenhuijzen} \& {de Jager}(1990)}]{nie90}
{Nieuwenhuijzen} H., {de Jager} C., 1990, \aap, 231, 134

\bibitem[{{Pols} \& {Marinus}(1994)}]{pol94}
{Pols} O.~R., {Marinus} M., 1994, \aap, 288, 475

\bibitem[{{Reimers}(1975)}]{rei75}
{Reimers} D., 1975, Memoires of the Societe Royale des Sciences de Liege, 8,
  369

\bibitem[{{Salpeter}(1955)}]{sal55}
{Salpeter} E.~E., 1955, \apj, 121, 161

\bibitem[{{Sansom} {et~al}\mbox{.}(2009){Sansom}, {Izzard}, \&
  {Ocvirk}}]{san09}
{Sansom} A.~E., {Izzard} R.~G., {Ocvirk} P., 2009, \mnras, 399, 1012

\bibitem[{{Scalo}(1986)}]{sca86}
{Scalo} J.~M., 1986, \fcp, 11, 1

\bibitem[{{Spurzem}(1999)}]{spu99}
{Spurzem} R., 1999, Journal of Computational and Applied Mathematics, 109, 407

\bibitem[{{Tout} \& {Eggleton}(1988)}]{tou88}
{Tout} C.~A., {Eggleton} P.~P., 1988, \mnras, 231, 823

\bibitem[{{Vassiliadis} \& {Wood}(1993)}]{vas93}
{Vassiliadis} E., {Wood} P.~R., 1993, \apj, 413, 641

\bibitem[{{V{\'a}zquez} \& {Leitherer}(2005)}]{vaz05}
{V{\'a}zquez} G.~A., {Leitherer} C., 2005, \apj, 621, 695

\bibitem[{{Zhang} {et~al}\mbox{.}(2004){Zhang}, {Han}, {Li}, \&
  {Hurley}}]{zha04}
{Zhang} F., {Han} Z., {Li} L., {Hurley} J.~R., 2004, \aap, 415, 117

\bibitem[{{Zhang} {et~al}\mbox{.}(2009){Zhang}, {Li}, \& {Han}}]{zha09}
{Zhang} F., {Li} L., {Han} Z., 2009, \mnras, 396, 276

\bibitem[{{Zhang} {et~al}\mbox{.}(2012{\natexlab{a}}){Zhang}, {Li}, {Zhang},
  {Kang}, \& {Han}}]{zha12}
{Zhang} F., {Li} L., {Zhang} Y., {Kang} X., {Han} Z., 2012{\natexlab{a}},
  \mnras, 421, 743

\bibitem[{{Zhang} {et~al}\mbox{.}(2012{\natexlab{b}}){Zhang}, {Han}, {Liu},
  {Zhang}, \& {Kang}}]{zhy12}
{Zhang} Y., {Han} Z., {Liu} J., {Zhang} F., {Kang} X., 2012{\natexlab{b}},
  \mnras, 421, 1678

\bibitem[{{Zhao} {et~al}\mbox{.}(2010){Zhao}, {Gao}, \& {Gu}}]{zhao10}
{Zhao} Y., {Gao} Y., {Gu} Q., 2010, \apj, 710, 663

\end{thebibliography}

\bsp
\label{lastpage}
\end{document}